\documentclass{aa}  

\usepackage{graphicx}
\usepackage{txfonts}
\usepackage{footnote}
\usepackage{tablefootnote}
\begin{document}

   \title{	
	The XMM deep survey in the CDF-S. X}

   \subtitle{X-ray variability of bright sources}

   \author{S. Falocco
          \inst{1}
          \and
          Paolillo M.
          \inst{2,3,4}
          \and
          Comastri A.
          \inst{5}
          \and
          Carrera F. J.
          \inst{6}
          \and
          Ranalli P.
          \inst{7}
\and
Iwasawa K.\inst{8,9}
\and
Georgantopoulos I.
\inst{10}
\and
Vignali C.
\inst{11}
\and
Gilli R.
\inst{5}
          }

   \institute{KTH, Department of Physics, and the Oscar Klein Centre, AlbaNova, SE-106-91 Stockholm, Sweden
\\
              \email{falocco@kth.se}
              \and
              University Federico II, Physics Department, Via Cintia, 80126 Naples, Italy
              \and
              INFN Napoli, via Cintia, 80126 Naples, Italy
              \and
              Agenzia Spaziale Italiana Science Data Center, via del Politecnico, 00133 Roma, Italy
              \and
              INAF Osservatorio Astronomico di Bologna, via Gobetti 93/3, 40129 Bologna, Italy
              \and
              Instituto de F\'isica de Cantabria (CSIC-UC), Avenida de los Castros, 39005 Santander, Spain
              \and
Lund Observatory, Department of Astronomy and Theoretical Physics, Lund University, Box 43, 22100 Lund, Sweden     
\and
              Institut de Ci\`encies del Cosmos (ICCUB), Universitat de Barcelona (IEEC-UB), Mart\'i i Franqu\`es, 1, 08028 Barcelona,
              Spain
              \and
              ICREA, Pg. Llu\'is Companys 23, 08010 Barcelona, Spain
              \and
              Institute of Astronomy and Astrophysics, National Observatory of Athens, Palaia Penteli, 15236 Athens, Greece
              \and
              University of Bologna, Department of Physics and Astronomy, via Gobetti 93/2, 40129 Bologna
             \thanks{}
             }

   \date{}

 
  \abstract
   {}
   {We aim to study the variability properties of bright hard X-ray selected Active Galactic Nuclei (AGN) in the redshift range between 0.3 and 1.6 detected in the Chandra Deep Field South (XMM-CDFS) by a long ($\sim$3 Ms) XMM observation.

   }
   {Taking advantage of the good count statistics in the XMM CDFS we search for flux and spectral variability using the hardness ratio techniques.
     We also investigated spectral variability of different spectral components (photon index of the powerlaw, column density of the local absorber and reflection intensity).
     The spectra were merged in six epochs (defined as adjacent observations) and in high and low flux states to understand whether the flux transitions are accompanied by spectral changes.}
   {The flux variability is significant in all the sources investigated. The hardness ratios in general are not as variable as the fluxes, in line with previous results on deep fields. Only one source displays a variable HR, anti-correlated with the flux (source 337). The spectral analysis in the available epochs confirms the steeper when brighter trend consistent with Comptonisation models only in this source at 99 \% confidence level. 
     Finding this trend in one out of seven unabsorbed sources is consistent, within the statistical limits, with the  15 \% of unabsorbed AGN in previous deep surveys.
   No significant variability in the column densities, nor in the Compton reflection component, has been detected across the epochs considered. The high and low states display in general different normalisations but consistent spectral properties.}
   {X-ray flux fluctuations are ubiquitous in AGN, though in some cases the data quality does not allow detecting them. In general, the significant flux variations are not associated with a spectral variability: photon index and column densities are not significantly variable in nine out of the ten AGN over long timescales (from 3 to 6.5 years). The photon index variability is found only in one source (which is steeper when brighter) out of seven unabsorbed AGN.  The percentage of spectrally variable objects is consistent, within the limited statistics of sources studied here, with previous deep samples. 

   }


   \maketitle
%

   \section{Introduction}

   Active Galactic Nuclei are characterised by strong variability at all wavelengths (e.g. \citealt{edelson1996}). Rapid X-ray variability is a distinctive property of AGN and provided the most compelling evidences of Super Massive Black Holes (SMBH) at their centres \citep{rees1984}.
 
 The observed variability can be purely in flux or also in spectral shape; while the former is the most common, it is also possible to observe changes in the spectra, sometimes correlated with the flux variations.
   
 Studies of the spectral variability of nearby AGN found that the AGN spectra are steeper when brighter (e.g. \citealt{nandra1997}), the softer primary power-law found when the source brightens is due to changes in the central engine \citep{zdziarski2003}, specifically to transitions between accretion disk states.
This trend has been confirmed in a sample of nearby AGN monitored regularly by the Rossi X-ray Timing Explorer  and explained by a model where a main power-law component, intrinsically variable in flux and in slope, is combined with a constant reflection component \citep{sobolewska2009}. These findings seem consistent with a scenario where the spectral changes are due to variations in the accretion rate: in the high accretion rate regime, the X-ray emitting corona is cooled more efficiently by the seed photons from the accretion disk, thus resulting in a higher overall flux but also a steeper power-law.

   Recent works have shifted from spectral variability to time lags between different spectral components. These studies found a delay between the relativistic reflection in the accretion disk and the primary continuum observed in the X-rays due to the light travel time from the corona to the reflecting accretion disk \citep{zoghbi2012}. 
 
 In some AGNs, the origin of the observed X-ray variability has been traced back to variations in the absorbing column density along the line of sight, occurring in the torus or in the Broad Line Region (BLR).
   This mechanism has been proposed to explain the X-ray variability of nearby AGN in \cite{risaliti2002}, \cite{malizia1997}, \cite{puccetti2004} and \cite{risaliti2010} where significant changes in the column densities of the X-ray absorber have been reported. This suggests that the absorbing torus has a clumpy structure and that the observed variability is partly due to orbiting clouds in the line of sight. The timescale of this variability goes from days to months.  A similar result has been found by \cite{torricelli2014} in a larger sample of AGNs, showing that occultations by the BLR may be the responsible for part of the observed variability above 2 keV over timescales of days.

Studies of the X-ray variability in high redshift AGNs have confirmed that flux variability is typical of these sources over months to years timescales \cite[e.g.][]{bauer2003,paolillo2004,papadakis2009,shemmer2014,lanzuisi2014,vagnetti2016}. The variability timescales explored so far have been extended to decades in \cite{middei2017}.  However, few of these works have been able to investigate in detail the spectral variability of their sources, mainly due to the low statistics:  spectral variability has been found in  the ensemble analyses presented by \cite{gibson2012} and \cite{serafinelli2017}.
\cite{paolillo2004}, studying the Chandra Deep Field South (CDFS), found evidence for intrinsic flux variability for the majority of the sample but spectral variability was observed in only $\sim30\%$ of the sources. Half of the spectrally variable sources show an anti-correlation between hardness ratio and flux, a behaviour compatible with the softer-when-brighter trend discussed above.  A similar result was found in \cite{mateos2007}, who found 80\% sources variable in flux in the Lockmann Hole and only 40 \% sources variable in X-ray colours at 3$\sigma$ level. Furthermore, \cite{shemmer2014} found a high redshift source (z=2.7) with significant spectral variability which they argued was due to a variable column density in the line of sight. More recently, \cite{yang2016} studied the CDFS dataset finding variability in X-ray luminosity in 74 \% of the AGN in the survey. They detect variability in column densities in only 16 \% of their AGN sample. 

   Taking advantage of the good count statistics in the XMM-Newton observation of the Chandra Deep Field South (XMM CDFS, \citealt{comastri2011}), we aim at exploring the flux and spectral variability in AGN checking the previously described pictures. For this purpose, we investigate the ten brightest sources in the XMM CDFS. The main purpose is to understand what is the main driver of the variability of high redshift sources, investigating the variability of several spectral components, i.e., photon index of the primary power-law, column densities, reflection.
The spectra have been analysed with Xspec version 12.9. The errors correspond to a 90\% confidence level for one interesting parameter ($\Delta\chi^2=2.71$) where not otherwise stated. The cosmological parameters are the default ones in Xspec: $H_0=70 \rm{km~s^{-1}~Mpc^{-1}}$, $\Omega_\Lambda=0.73$, $\Omega_{tot}=1$ \citep{komatsu2011}.


\section{Source sample and data analysis strategy}
\subsection{The data}
This work is based on an observation of the Chandra Deep Field South with XMM-Newton for a total of 3.3 Ms of full exposure time (hereafter, the sample is called XMM CDFS). The cleaned exposure times from the EPIC pn and MOS cameras are 2.5 Ms and 2.8 Ms respectively. The XMM CDFS has 33 observations obtained in two distinct periods, i.e. 2001-2002 and 2008-2009. The details on the data reduction and the source detection are presented in \cite{ranalli2013}. 

Despite the larger sensitivity of the EPIC pn detector, a comparison of the background levels of the MOS and pn \cite[see][]{ranalli2013} shows that the background of the MOS camera is always lower and more stable than the pn. In addition, the number of sources falling in the gaps of pn CCDs is larger than for the MOS detectors, due to their different geometry, thus making more challenging to extract consistent lightcurves and spectra over several years for sources close to CCD gaps where the background contribution is hard to estimate properly \citep{ranalli2013}.  We thus choose to limit our analysis to the MOS data.

The 33 observations are grouped in the six main intervals: July 2001, January 2002, July 2008, January 2009, July 2009, January-February 2010 (see Table 3 of \citealt{iwasawa2015}). We note that there are inhomogeneities in the lightcurve sampling because the dead MOS1 CCD \citep{ranalli2013} affects the sampling of sources in different positions of the field-of-view due to the variable aim point of the observations and roll angles.  The background level of the observations has increased over time, being lower in the 2001-2002 observations than at later epochs  \citep{ranalli2013} while, on the other hand the first epoch is composed by less observations than the following ones, and thus has a smaller overall exposure time. These effects raise the uncertainties on the spectral parameters of the six epochs.

\subsection{The sample}
This work focuses on the ten AGN in the 2-10 keV catalogue published by \cite{ranalli2013} defined as having more than 5000 time-integrated counts in the source detection band, 2-10 keV observed frame frame band.
The ten AGN have spectroscopic redshifts from 0.3 to 1.9 (from \citealt{treister2009,silverman2010,szocoly2004} and \citealt{luo2016}) and five of them have an optical spectroscopical classification from the literature. Sources with id. 319, 337 and 203 (the identification numbers refer to the catalogue of \citealt{ranalli2013}) are classified as Broad Line AGN by \cite{szocoly2004}.
Source 33 is classified as unabsorbed AGN in the optical spectroscopic study presented in \cite{treister2009}; source 48 is classified as a narrow emission line galaxy in \cite{silverman2010}. 

\begin{figure*}
 \centering
  \includegraphics[width=16cm]{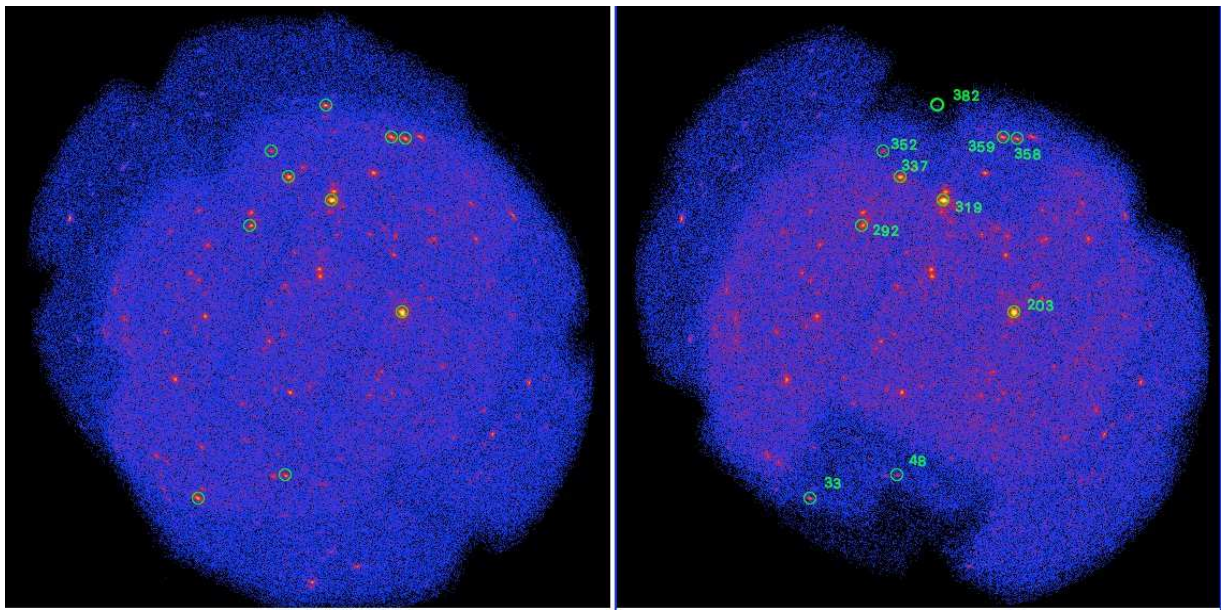}
  \caption{Spatial distribution of the sources investigated in this study overlaying on the stacked MOS1 (right hand panel)/MOS2 (left hand panel) field of view (in the 0.5-8 keV band).}
          \label{field}
    \end{figure*}

The variability of sources 203 and 319 has been studied previously in \cite{iwasawa2015}, but here we include them because we use a different approach, and also to check the consistency of our results;  sources 33 and 48 have been described as absorbed sources from the properties of their integrated spectra in \cite{iwasawa2012}  and in \cite{georgantopoulos2013} respectively; source 352 has been studied by \cite{vignali2015} who found features from an outflowing wind in the integrated spectrum.
Table \ref{Tabproperties} summarises the properties of the 10 AGN. The position of the sources within MOS1 and MOS2 field-of-view are displayed in Fig. \ref{field}.

 \subsection{Analysis strategy}
In this context our work explores the spectral variability of the CDFS sources using an approach that combines lightcurves, hardness-ratios and spectral analysis, investigating both individual epochs as well as the combined spectra of the high and low states (see below), in order to constrain the origin of the flux variability and its connection with the primary emission mechanisms.

As a starting point we extracted the flux lightcurves of each source separately for the MOS1 and MOS2 detectors, in the 0.5 - 8 keV observed frame band, binning the data in individual observations.
The lightcurves are shown in the top panels of figures 2-12, where it can be noted that the data points are, on average, consistent between the MOS1 and MOS2. The points where the measured fluxes differ significantly are usually due to source positions at the edge of the FOV or close to the chip gaps where the exposure corrections are more uncertain. We thus compute for each epoch the average of the fluxes from the two cameras (the average is weighted with the errors), as the best estimate of the intrinsic single-epoch flux.
 The six panels in the Figures refer to the six epochs mentioned earlier and defined in \cite{iwasawa2015}.
We assess if a source is variable based on the comparison with 1000 simulations of each lightcurve in each detector \citep{paolillo2004}; we assume a constant flux set to the average source and background flux of the real source.
We characterise the source variability through the normalised excess variance (NXV) for each lightcurve \citep{nandra1997,turner1999,edelson2002}. We then take into account the different lenghts of the lightcurves in their rest-frame applying the correction for redshifts as explained in \cite{vagnetti2016}.
This estimate allows a comparison of variability amplitudes between the ten sources analysed here, and with known samples in the literature.

Hardness Ratios (HR)\footnote{Hardness Ratio is defined as $HR=\frac{F_{2-8}-F_{0.5-2}}{F_{0.5-2}+F_{2-8}}$ where $F_{0.5-2}$ and $F_{2-8}$ are the fluxes in the 0.5-2 keV and 2-8 keV bands.} lightcurves are then constructed to investigate colour changes of the sources over the timescales explored here. The HR lightcurves are derived with the same sampling as the flux lightcurves and serve as a preliminary spectral variability assessment, to be  refined through a proper spectral analysis when the statistics is sufficient.
The bands are in the observed frame, so for the highest redshift sources the HR is not actually sensitive to the softest spectral features (i.e. to the $\rm N_H$). For each source we further plot the HR vs flux to identify correlations between the flux and spectral variability.
 

In addition to the flux and HR lightcurves, we performed a more detailed spectral analysis.  For this purpose, the spectra of the individual epochs have been merged by grouping them in two different ways: the spectra have been first grouped in the six time intervals as in \cite{iwasawa2015}, corresponding to the panels shown in the top row of Figures 2-12, to check if any variability of the spectral features is related to the sampled timescales. Subsequently, they have been grouped in high and low states with the main purpose to understand whether the flux changes correlate with spectral changes as well. The high and low flux states of MOS1 and MOS2 are defined as having flux higher and lower than the average flux level over the entire lightcurve (black dotted line in Figs. 2-12, top panel). 

The merging of the MOS1 and MOS2 spectra corresponding to the same source, observation and filter, was done computing the straight sum of the counts and the exposures. The backscale of the total spectrum was computed as the average of the backscale values of the individual spectra weighted for the contribution of each exposure. The response matrix and ancillary file of the combined spectrum have been constructed in the same way, as the weighted average (computed using ftools task addrmf). The procedure just described was repeated for the merged spectra of the six epochs and of the high and low flux states.
The background-subtracted spectra of each epoch and of the high and low states spectra were rebinned in order to have more than 20 counts in each bin for the $\chi^2$ statistics to be valid. The final spectra of the ten sources have different quality in the six epochs, ranging from $\sim$100 net counts in the worst case (the lowest counts spectrum of source 352) to $\sim$15000 net counts for the best case (the highest counts spectrum 203) in the band used for the spectral analysis detailed below (0.5-7 keV rest-frame band).




\begin{table*}
  \caption[]{Source properties: (1): source IDs, (2): coordinates,  (3) spectroscopic redshifts, (4): net integrated counts in 0.5-7 keV rest-frame bands from MOS1 and MOS2 instruments, (5) Variability probability combined from MOS1 and MOS2 ($1-[(1-P_{MOS1})\times(1-P_{MOS2})]$), (6) excess variance (weighed over the errors) of MOS1 and MOS2, (7): excess variance corrected for redshifts,
    (8): continuum luminosity in the 0.5-7 keV rest-frame band (average between the high and low state spectra, from the best fits in Table 2); (9) fractional flux variability from the lightcurves; (10): fractional variability in the powerlaw normalisations in the spectra of the available epochs}
      \label{Tabproperties}
      \centering
         \begin{tabular}{r c c c c c c c c c}       
            \hline
            Source      & coord 
            & z  &   C & $P$   &  NXV  & $NXV_c$  & L  &  $f_{LC}$    &  $f_{epochs}$  \\
                  &  &   &   $10^3$ &   &   $10^{-2}$ & $10^{-2}$ &    $10^{44}$erg/s  & \% & \%  \\
(1) & (2) & (3) & (4)  & (5)  & (6) & (7) & (8) & (9)  & (10) \\
            \hline
            203 &  03:32:08.62  -27:47:34.80  
            & 0.544  &  46.0 & 1  & $2.55\pm0.64$     &    2.83 &  $1.30\pm0.03$   & 17  & 11 \\ 
            319 &    03:32:26.95 -27:41:06.00 
            & 0.742    & 35.5 & 1  & $3.07\pm0.62$ &    3.51    &  $3.24\pm0.04$ &    19   & 13  \\ 
            358  &  03:32:07.85 -27:37:32.16 
            & 0.976    &   17.8 & 1 & $2.99\pm1.58$ &  3.52   & $1.4\pm0.5$  &   19   & 10  \\ 
            337 &  03:32:38.14 -27:39:44.64  
            & 0.837    &   11.2  & 1    & $3.77\pm0.77$ &  4.36  & $1.67\pm0.04$ &  21   & 23 \\ 
           
  359 &  03:32:11.54 -27:37:27.84  
  &   1.574   &  6.9  & 1 &  $4.15\pm0.94$ &   5.20  &  $5.35\pm0.18$ &  23 & 18 \\ 

        292 & 03:32:48.05 -27:42:33.12   
            &   0.979   &  5.5   & 1  & $3.46\pm0.96$ &  4.08  & $0.91\pm0.15$ &     20  & 15 \\ 
    33 &   03:33:01.70 -27:58:18.84    
            &  1.843   & 4.0 & 1  & $2.38\pm0.92$ &   3.06  & $9.2\pm0.3$   &  18  &  18 \\ 

    382 & 03:32:28.32 -27:35:35.88     
            & 0.527   & 2.5  & 1  & $12.7\pm3.7$ &   14.06 & $0.59\pm0.06$  & 38  & 28\\ 

    48  &  03:32:38.93 -27:57:00.72   
            &  0.298  & 2.3 & 1   &  $1.51\pm0.49$  &  1.61  & $0.07\pm0.03$ &  13   & 16   \\ 
        352 &   03:32:42.46 -27:38:15.72    
            & 1.6   & 1.5 & 0.99   &  $2.90\pm1.81$  &    3.65   & $1.23\pm0.60$  &   19  & 12 \\ 
            \hline
         \end{tabular}
   \end{table*}

 \subsection{Modelling the X-ray spectra}

After assessing if we detect significant flux or spectral variability, we aim at checking if the observed spectral changes are due to, e.g., changes in the absorbing column due to the circum-nuclear material in the line of sight, or to changing conditions of the inner central engine. 

The fits have been performed in the 0.5-7 keV rest-frame band starting with a 'baseline model' and adding more components when needed.
Since our sources span a broad redshift range up to z $\sim$ 1.9, the cut at 0.5 keV rest-frame energy  represents a compromise between the need to include the soft X-rays in order to assess the presence of a soft excess component, and the attempt to avoid the observed-frame energies below 0.2 keV potentially affected by the background. On the other hand, the high energy cut at 7 keV rest-frame energy has been set to exclude the spectral region mostly dominated by the background.

The starting 'baseline model' is composed by an absorbed power-law plus a gaussian emission line.
We used fixed Galactic absorption with $N_H=8\times10^{19}\rm{cm^{-2}}$ (represented by the wabs model in xspec) and free local absorption at the redshift of each source (with the model zwabs in xspec).
In the baseline model the free parameters are the normalisation of the powerlaw, its slope and the column density of the local absorber. In this model we assume that there is negligible reflection (Compton scattering) and negligible soft excess.  
When the goodness of fit was not acceptable on the basis of the reduced $\chi^2$, we added other components or let specific parameters free. 
 The gaussian in the baseline model has central energy fixed at 6.4 keV and width 0.1 keV: we let the centroid energy free where the data seem to require a slightly different energy (as explained in section 3).

Three spectra (those of sources 358, 292 and 48) in this study required an additional component to account for the excess observed at the spectral energies below 2 keV. 
The mekal model in xspec is the first one used to fit the soft excess, it represents thermal emission (Bremstrahlung and lines) which produces lines in the soft-X-rays. This thermal component is emitted from ionised material far from the central engine. This collisional model is good in first approximation for the Narrow Line Region (NLR) even if it is photoionised.  
The fixed parameters adopted for the mekal model are the hydrogen density $\rm{n_H=1\times cm^{-3}}$ and the metal abundance=1 (Solar abundance). The temperature is free to vary during the fitting. 
A second attempt to fit the soft excess has been made considering Compton up-scattering of the photons from the disk, represented by an additional powerlaw component added to the baseline model. We fixed its slope to that of the primary powerlaw and kept the normalisation as a free parameter. 
 
 For two more sources (48, 352) the data suggest the presence of a Compton reflection component. We model this contribution through the \textit{pexrav} model, set up to represent the reflected component only (the direct one is already accounted for by the power-law of the baseline model).  The values of the abundances have been fixed to Solar values, the angle to $45^{\circ}$, the cutoff energy to 100 keV. Since the slope of the reflected component should be the same as the primary power-law, we fixed their photon indices to the same value.

  \begin{table*}
    \caption[]{Fits results. Col 1: Source ID; Col 2: fitting model in the 0.5-7. keV rest-frame band: baseline (absorbed powerlaw and a gaussian), E free (baseline with Energy of gaussian free), mek (baseline+mekal) to account for the soft excess, pex (baseline+pexrav), pex E free (baseline+pex with centroid of the gaussian free), pow: baseline+pow to account for the soft excess; Col 3: column density of local absorber in $10^{22}\rm{cm^{-2}}$; Col 4: Photon index of the observed powerlaw;
      Col. 5: EW of the narrow Fe line at 6.4 keV expressed in eV; Col 6: temperature of  material emitting the mekal component (mek). Col. 6 represents the centroid energy of the iron line when left free and the reflection fraction of the Compton reflection when added. Col 7: chi square / d.o.f. Col 8: linear correlation  coefficients obtained with the least-squares regression between photon index and the flux (obtained from the spectral fits of the six epochs in the band between 0.5-7. keV). Col. 9: p-value for the least-squares regressions reported in col. 8: these represent the probability that the two parameters are not correlated. In Col 8 and 9, the least-squares regressions have not been computed for sourcer 382 (because it has only the spectra of two epochs).
      }
         \label{Tabfitsresults}
         \begin{tabular}{c c c  c c  c  c c c  }

 (1) & (2) &   (3)  &   (4)   &  (5)  & (6)  & (7) & (8) & (9)  \\         
           source & fit & $\rm N_H$  & $\Gamma$ 
           & EW  & T, E, R  & $\chi^2/dof$  & $r$   & p \\
   &  & $10^{22}\rm{cm^{-2}}$  &  
           & eV  & KeV, KeV, -  &   &    &  \\
           
\hline \hline
$203_h$ & \emph{baseline}  & $5.5\pm1.5\times10^{-2}$ & 1.97$\pm0.04$ 
& $<91$ &- & 244.06/239  
&          -0.412 & 0.42 \\
$203_l$ & \emph{baseline}  & $5.8\pm1.3\times10^{-2}$ & 1.94$\pm$0.04 
& 159$\pm53$ & - &303.20/267 &  \\

\hline \hline
$319_h$ & \emph{baseline}  & $<$0.007 & 2.04$\pm0.02$  & 158$\pm$52 & -  &  215.00/207
&          -0.06  & 0.91  \\ 
$319_l$ & \emph{baseline}  & $<$0.004 & 2.06$\pm$0.02  & 169$\pm$56 & -& 181.19/197 &  \\

\hline\hline
$358_h$ & \emph{baseline}  & 7.2$\pm1.8$ & 1.35$\pm$0.45  & $<101$ & - &   67.01/77
&         -0.491  & 0.67  \\

$358_h$ & E free & 8$\pm2$ & 1.46$\pm0.50$ & $<138$ & E= $3.37\pm0.27$ & 66.02/76  \\
$358_h$ &  mek & 7.7$\pm2.0$ & 1.4$\pm$0.5   &$<109$ & T $<0.23$ &57.79/75 &  \\
$358_h$ &  pow & 9.2$\pm2.0$ & 1.68$\pm0.47$  &$<125$ & $-$  &58.68/76 &  \\
$358_l$ & \emph{baseline}  & 9.7$\pm2.2$ & 1.91$\pm$0.55  & $<136$ & - & 49.27/59 & \\
$358_l$ &  mek & 10.7$\pm$2.6  & 2.1$\pm$0.6  & $<152$ & T $<0.91$ &39.64/57 &  \\
$358_l$ &  pow & 11.3$\pm2.1$ & 2.18$\pm0.53$  &$<156$ & $$  & 41.90/58 &  \\
\hline \hline
$337_h$ & \emph{baseline}  &  $<$0.01& 2.23$\pm$0.04&  172$\pm$91 & - &   139.55/152
&        0.986 & 0.01  \\


$337_h$ & E free  &  $<0.01$  & $2.23\pm0.04$ &   239$\pm$100 & E= 3.59$\pm$0.07 & 133.91/151 &  \\

$337_l$ & \emph{baseline}  &  $<$0.05  & 2.18$\pm$0.07 &   310$\pm$159 & - & 107.66/112 &  \\


\hline\hline
$359_h$ & \emph{baseline}  & $<0.05$ & $1.70\pm0.08$  & $<80$ & -   &  121.24/108
&        -0.907 & 0.28   \\
$359_l$ & \emph{baseline}  & $<0.06$ & $1.69\pm0.09$  & $<97$ & - & 119.52/108 &  \\

\hline \hline
$292_h$ & \emph{baseline}  & 0.78$\pm$0.20 & 1.40$\pm$0.16 
& $<84$ & - &  140.00/135   
&             0.26 & 0.67 \\
$292_h$ &  mek & 1.04$\pm$0.30 & 1.54$\pm$0.20  & $<$108 & T= 0.219$\pm0.090$ &130.63/133 &  \\
$292_h$ & pow  & 1.49$\pm0.50$ & 1.64$\pm0.24$  & $<124$ & $$ &  132.32/134 &  \\

$292_l$ & \emph{baseline}  & 1.1$\pm0.3$ & 1.58$\pm$0.22 
& $<$141 & - &  101.01/104  &  \\


\hline \hline
$33_h$ & \emph{baseline}  & $<0.04 $ &  1.80$\pm$0.07  &  $<138$ & -  &    110.20/116    
&         -0.99 & 0.08    \\
$33_h$ & Efree  & $<0.05$  &  1.82$\pm0.08$  &  93$\pm60$  & E= 2.1$\pm0.1$ & 106.68/115 &  \\
$33_l$ & \emph{baseline}  & $<0.14$  &  1.87$\pm$0.10  &  155$\pm92$  & - & 83.91/101 &  \\
$33_l$ & Efree  & $<0.13 $  &  $1.87\pm0.20$  &  $190^{+800}_{-110}$  & E= $2.31\pm0.20$ & 83.09/100 &  \\


\hline\hline

$382_h$ & \emph{baseline}  & 0.15$\pm0.06$ & 1.64$\pm0.13$  & $<226$ & - & 70.76/74   &    -               \\
$382_l$ & \emph{baseline}  & 0.19$\pm0.10 $ & $1.66_{-0.18}^{+0.20}$  & $<187 $ & - &24.11/39  & -  \\

\hline\hline
$48_h$ & \emph{baseline}  & 2.32$\pm0.60$ & 1.41$\pm0.32$  & 158$\pm150$ & - &  82.93/70    
&         0.813  & 0.39\\

$48_h$ & mek & 2.6$\pm0.6$ & 1.48$\pm0.32$   & 172$\pm142$ & T= 0.71$\pm0.20$ &  66.9/68 &   \\


$48_h$ & pow & 3.2$\pm0.8$ & 1.62$\pm0.35$  & 194$\pm150$ &  & 72.33/69 &   \\

$48_h$ & pex  & 4.3$\pm1.0$ & 2.48$\pm0.30$  & 144$\pm140$ &R$>-1$ &71.85/68 &   \\
$48_h$ & pex Efree & 4.3$\pm1.0$ & 2.48$\pm0.30$ & 148$\pm138$ & R$>-0.4$, E=4.89$\pm0.20$ & 71.54/67 \\

$48_l$ & \emph{baseline}  & 1.5$\pm0.9$ & 1.03$\pm0.50$  & $<211$ & - & 60.07/44 &   \\

$48_l$ & mek & 3.6$\pm2.0$  & 1.5$\pm0.7$  & $<275$ & T$<6 $ & 47.92/42 &   \\


$48_l$ & pow & 3.7$\pm1.5$ & 1.59$\pm0.54$  &$ <304$ &  & 47.61/43 &   \\

$48_l$ & pex & 3.7$\pm1.5$ & 1.59$\pm1.00 $  & $<303$ & R$>-0.25 $ &  47.56/42 &  \\
$48_l$ & pex Efree & 3.7$\pm1.4$ & 1.6$\pm1.6 $ & $92^{+300}_{-30} $ & R$>-0.25$, E= $4.56^{+0.80}_{-0.04} $ & 43.4/41  \\

\hline\hline
$352_h$ & \emph{baseline}  & 8.2$\pm3.5$ & $<1.1$  &$<180$ & - &  72.20/34   
&         -0.924 & 0.25   \\

$352_h$ & pex  & 25$\pm6 $  &  1.72$\pm0.20$  &  $<178$  & R $>-0.25$ & 41.18/32 &  \\
$352_l$ & \emph{baseline}  & $<2.6$ & $<0.7$  & 419$\pm200$ & - &31.77/26 &  \\
$352_l$ & pex  & 51$\pm10 $  &  $<1.4$  &  $<1000$  & R $>-1$ &   18.77/24 \\

    \end{tabular}
   \end{table*}

  \section{Results and discussion}

We investigated the spectra of the sources to get more detailed information than the HR study.
Some studies propose an interpretation of the AGN X-ray variability in scenarios similar to Galactic Black Holes but scaled by the BH masses, so it is characterised by larger timescales in AGN (from months to years, following \citealt{mchardy2006}).  In particular, \cite{papadakis2009} and \cite{sobolewska2009} found a correlation between the observed flux variability and the spectral properties in AGN. In these works they reported  an intrinsic difference in the powerlaw slopes due to a varying accretion rate in each source. In fact, the powerlaw is produced through inverse Compton scattering by a corona close to the disk. The dissipation of the corona strongly depends on the geometry of the system and it is inversely proportional to the accretion rate $\dot{m}$ in the geometry described by \cite{esin1997}.
 In this scenario, the efficiency of the coronal cooling will be proportional to $\dot{m}$ as well as to the slope of the primary powerlaw in the X-rays. 
 As proposed by \cite{esin1997}, disk transitions from 'low' to 'high' states are the reason for the observed variability in the X-ray binaries.
As $\dot{m}$ increases, the inner radius of the accretion disk approaches the SMBH, the primary radiation (soft X-rays) impinging to the corona is higher, the coronal cooling is more efficient and the X-ray primary continuum spectrum will be steeper as described in \cite{zdziarski1999}.
 In the 'high' state, the disk is extended down to regions very close to the SMBH and $\dot{m}>1$.
 The term 'low' state, instead, refers to a  truncated disk and $\dot{m}<1$. In this case, the thermal emission from the accretion disk will be lower, the corona cooling will be less efficient and the power-law continuum will be harder \citep{zdziarski1999}.  This kind of variability can occur over a variety of timescales, from minutes to months \citep{belloni1997,fender2004} for the binaries; the variability timescales are scaled with the BH masses and luminosities following \cite{mchardy2006}'s empirical relation, so they are typically longer for AGN (from months to years) and detectable in the monitoring baseline of the XMM CDFS.

\subsection{Lightcurves in fluxes and in Hardness Ratios}

Our photometric analysis shows that all the sources investigated here show significant flux variability with a confidence level $\geq 99\%$ (Table \ref{Tabproperties}). The lowest probability is found for the faintest source (i.e. source 352), in agreement with the expectation that all AGNs are intrinsically variable and only the data quality prevents us to detect variability at low fluxes \cite[see, e.g.][]{paolillo2004, young2012, yang2016}.
The measured excess variance NXV spans values between 0.01 and 0.1, corresponding to flux variations in the range 13--38 \% over timescales between three and 6.5 years (at rest-frame), in agreement with what is observed in previous studies of the CDFS (\citealt{paolillo2004}; \citealt{paolillo2007}; \citealt{paolillo2017}).
We did not find any strong anti-correlation between variability (NXV) and the X-ray luminosity (found for example in \citealt{almaini2000,paolillo2004,lanzuisi2014,yang2016}); however this is not surprising because we probe a relatively narrow range of luminosities and also because, in order to measure such trend, we would need multiple lightcurves of the same source or a larger sample of sources with similar properties \citep{allevato2013,vaughan2003}.


  The HR lightcurves show that the flux variability does not correspond to a significant X-ray colour variability because the HR do not vary across the individual epochs in the majority of the sources (see second row of Figs. from 2 to 12). To understand this result,  we first underline that the sensitivity to the HR variability is strongly dependent on the data quality, since the limited counts statistics may prevent us to detect it (as pointed out also in \citealt{paolillo2004} and \citealt{mateos2007}).  This explains, at least in part, why the HR variability is common in nearby AGN characterised by spectra with good counts statistics \citep{papadakis2009}.
  On the other hand, a larger sample size would certainly increase the chance to find spectrally variable sources, e.g. the percentage was 30 \% in the larger sample constructed from the previous CDFS by \cite{paolillo2004}.
  HR variability has been also found in 20$\pm6$\% of the 123 brightest AGNs in the LH field \citep{mateos2007}. \cite{mateos2007} point out that this fraction increases with the quality of data, and can potentially reach up to 40\%.
  
In the Comptonization scenario discussed above, the HR is expected to anticorrelate with flux, since in the bright states the seed photons cool the corona thus softening the spectrum. We looked for evidence of such behaviour in our sample, but a large scatter in the HR versus flux trend and the large uncertainties on the two quantities prevent us to clearly detect such trend. Having said that, a marginal decrease of the HR with the flux can be barely seen in sources 337, 382 and 33. 
Only for source 337, a spectral steepening with the increasing flux is confirmed by the spectral analysis described below. On the other hand, a clear evidence for a spectral hardening with the increasing flux is not found in any of our sources.
Our results are generally in agreement with the finding that this sample does not display a significant spectral variability for the majority of the sources, except for source 337 that instead agrees with the 'steeper when britghter trend'. 
While it is possible that the limited spectral variability observed here is a consequence of the limited spectral quality in the individual observations and of the limited sample size, it is also possible that the changes in the spectral components are more complex than a simple HR analysis can reveal on a epoch-to-epoch basis. We thus investigate the sources in more detail through temporal-resolved spectral analysis.
We point out that the large HR found for sources 358, 48, 352 already suggest the presence of a hard (reflection) component, which is confirmed by the improved spectral analysis discussed below.

\subsection{Time-resolved spectral analysis}

 The single-epoch spectra have been first fitted with the baseline model to look for variations in the spectral components, focusing on the power-law normalisation and slope, amount of absorption and iron line properties.
 We first assessed if there is variability in the power-law normalisations, finding that, across the six epochs, it goes from 10 \% from the source where it is less variable (358) to 28 \% for source 382, as can be seen in Table 1.
 In general, the power-law slopes variability is not as strong as the flux variability, since the power-law slopes vary at most by 15\% across the six epochs (in sources 203 and 337).
 To check whether the results found for local AGN  can be extended to our sample, we explored the trends between power-law slopes and continuum fluxes that in \cite{sobolewska2009} are anticorrelated.
 We measured the linear correlation coefficients between the power-law photon index $\Gamma$ and the continuum flux, 
 as well as the probabilities for the correlations, using the least squares regression. Table 2 (col 8) shows the linear correlation coefficients (r) and the p-values (which represent the probabilities that the two measurements are not correlated, see col 9) obtained with the linear least-squares regression between photon indices and fluxes for the epochs where measurements of the two parameters could be derived from spectral fitting. The correlation coefficients reported in the table do not take into account the uncertainties on the measurements.
  Source 337 is steeper-when-brighter, confirming the trend anticipated by the HR versus flux discussed in the previous section.
 For source 203 and 319 we did not find the 'steeper-when-brighter' trend, in agreement with \cite{iwasawa2015}. 
 For the remaining sources we did not find a clear evidence for a trend between photon index and flux; for source 382 we could not do the linear correlation due to the large uncertainties on the two measurements.

In short, the power-law slope $\Gamma$ does not appear to be significantly variable for the majority of the sources, as opposed to the fluxes of the primary power-laws.To test if this result is affected by changes in the absorbing column along the line of sight we performed additional fits to the spectra in the 2-7 keV rest-frame band and confirmed that there is no significant variability in the power-law slopes, consistently with the fits in the 0.5-7 keV rest-frame band.

We tried to determine whether the absorber column density presents variability in the epochs considered but the values found for the column densities in the individual epochs are broadly compatible within the uncertainties. We also tried to study the iron line variability and we found that the EW are consistent within the uncertainties across the epochs explored in this survey.

\subsection{Flux-resolved spectral analysis}
Since the individual epochs do not display any evident spectral variability we merged the spectra of high and low flux states, defined in Sect. 2.
The high and low state fluxes differ, on average, by about 10\% in our sample.
The powerlaw slopes are broadly consistent within the uncertainties in the high and low flux states spectral fitting.
We also repeated the spectral analysis of the sources in the 2-7 keV restframe band in order to check the variability of the power-law slope ignoring the spectral region most affected by photoelectric absorption. This analysis confirmed that the general lack of change of photon index across the high and low states in the majority of the sources is not due to local absorbers. The 'steeper-when-brighter trend' found for source 337 in the time-resolved spectral analysis is reflected in the flux-resolved spectral analysis as a slightly steeper spectrum in the high flux state, albeit the two values are consistent within the uncertainties (see Col. 4 of Table 2).

For those sources where we observed an excess with respect to the baseline model in the soft X-rays, we have introduced additional spectral components to model such emission. Several processes can contribute to the observed soft X-rays. One possibility is that the soft excess is due to Compton up-scattering of the disk photons as described in \cite{done2012}. The NLR can also produce radiation in the soft X-rays, as a thermal emission (that we can model through a mekal component in xspec as was made in \citealt{corral2008}). In some nearby AGN it is often possible to detect a significant host galaxy emission, also fitted by a mekal, which may contribute to the observed soft X-ray emission (e.g. \citealt{hernandezgarcia2016}). 
In our sample, the soft excess is required to fit the spectra of three sources (292, 358, 48) and it has been equally well modelled by the thermal emission  and by a scattering component. For source 292 this feature is detected in the high state only, perhaps because of its better counts statistics (adding such component in the fit of the low state spectrum does not improve the $\chi^2$). In sources 358 and 48 it is detected in both high and low flux states.

The high and low flux states allowed to determine the EW of the narrow iron line that was unconstrained in the individual epochs due to the insufficient statistics.
A decreasing EW with an increasing luminosity, found for the first time in \cite{iwasawa1993}, has been explained considering that an active nucleus might blow material (where the iron line originates) away through thermal dissipation or a strong radiation pressure.
Narrow iron lines have been searched for the spectra merged in the high and low flux states to check whether the EW is lower in high flux states. 
 In sources 319 and 337 they are compatible across the high and low state. In a few other sources (203, 33 and 352) instead, the lines are detected only in the low flux state because the EW uncertainties (90\% confidence level) exclude the zero value. In the same sources the high flux states can only give upper limits to the EW. Since the high state spectra are characterised by typically better statistics, the lack of the line feature is not due to an insufficient spectral quality but it is an intrinsic property of those spectra and may be explained as line dilution by the intrinsic continuum in the high flux states. 
 For the remaining sources (292, 358, 359, 48, 382) the line could not be detected. 

 The addition of a relativistic iron line, expected in the accretion disk scenario, to the narrow line just discussed, is not required to fit the data (in all the sources) because a better spectral quality is needed to reduce the uncertainty on the continuum below the line.

 The X-ray variability has provided interesting clues not only on the central engine of AGN but also on their circum-nuclear material.
For instance \citet{risaliti2002,risaliti2010,torricelli2014} found variations in the absorber column densities in nearby AGNs, when analysing the photoelectric absorption features in their spectra.  This has helped to assess the morphological properties of the absorber, supporting its clumpy structure \citep[e.g][]{risaliti2010} and constraining the recurrence cycle of BLR eclipses \citep{torricelli-ciamponi2014}.
 According to the X-ray spectral analysis, only three sources in our
 sample (358, 48, 352) are moderately absorbed with column density in
 the same range covered by \cite{torricelli2014} (1-7$\times10^{23}\rm{cm^{-2}}$), while he rest of the sources do not display any significant
 absorption. Source 352 displays some evidence for an $\rm N_H$ change between the high and low state (see Table 2) but the large uncertainties on the fit parameters ($\rm N_H$ and power-law photon index) do not allow to draw any robust conclusion on its $\rm N_H$ variability.
 Amongst the sources where we do detect absorption
 we do not find reliable evidence for $\rm N_H$ variations between high and low states nor in the spectra of the individual epochs. While this can be an effect of the limited statistics and of the insufficient spectral quality, it is also possible that the timescales explored here do not allow to observe such variability. Indeed, our months-years timescales are longer than
 those of \cite{torricelli2014} (which goes from minutes to months), and this may prevent the detection of fast $\rm N_H$ changes detected in those studies.
Moreover,
 it is not surprising that our sample lacks the $\rm N_H$
 variations that were detected in \cite{torricelli2014}, since we have
 only three candidate absorbed sources, so our sample is biased toward bright, unabsorbed 
 AGNs.
 
In few strongly absorbed AGN observed by Chandra it has been possible to detect  column density variability:
 \cite{yang2016} studied the 68 brightest sources in the 6 Ms CDFS and found one strongly absorbed AGN with remarkable
 variability in the absorber (J033218.3-275055 in the Chandra sample). The source, with
 column density of 3$\times10^{23}\rm cm^{-2}$ and photon index of 1.2,
 is classified as a Compton Thick.
 \cite{yang2016} also found a source with transitions from an obscured to an unobscured state interpreted as an eclipse event (J033229.9–274530 in that sample).
 In \cite{yang2016}, two more sources are found to be variable  (J033226.5-274035, J033259.7-274626 in that sample), but their spectral variability is less strong than the flux variability, which is instead very significant across the Chandra epochs.
 These  sources are not included within
 the ten AGN with the largest number of counts in the XMM-CDFS investigated in the present work. 

 We have explored the possibility of detecting
 the Compton reflection component in the spectra of the three sources
 with X-ray absorption (358, 352 and 48). While for source 358 the
 Compton reflection component is not required by the data (it does not improve the chi square), it is required in sources 352 and 48 (see the chi square and the degrees of freedom in column 7 of Table 2), in agreement with their HR values, ranging from 0.5 and 1 (see Fig. 10 and 11).
 The reflection component is not variable, consistently with
 emission from material located far away from the central SMBH. Low redshift AGN with high quality spectra show similar results, for instance a constant Compton reflection with a highly variable primary
 powerlaw was found in the nucleus of the Seyfert 1 NGC4051 \citep{lamer2003}.

%

 \section{Summary and conclusions}

We studied the X-ray spectral and flux-variability of a sample of AGN detected in the XMM observations of the CDFS.
We assessed the flux variability of the sources constructing the MOS1 and MOS2 lightcurves. We estimated the excess variance and the variability probability using montecarlo simulations. 
A preliminary assessment of the spectral variability of each source was done examining the single-epoch HR in each epoch  and looking for correlation between HR and flux.  
We then performed a more refined spectral analysis merging the spectra in six epochs as well as merging the spectra of high and low flux states. The analysis allowed us to check if the detected variability is due to intrinsic changes in the normalisation of the powerlaw continuum, in its slope or to changes in the column densities of the absorber along the line of sight. We additionally explored whether we could detect the  6.4 keV iron line and if it was variable as well. 
We then assessed the presence (and variability) of any soft excess and we measured the contribution of a possible Compton reflection component.

Our main result is that most of variability is due to intrinsic flux changes, related to continuum normalisation changes, but the main spectral components (e.g. power-law continuum slopes, $N_H$, iron line equivalent width) are mostly constant or marginally variable.
This confirms previous results on AGN up to redshift $\sim3.5$, in several deep fields, from the first observations of the CDFS studied by \cite{paolillo2004} to the most recent observations of the same field by \cite{paolillo2017}, as well as the LH field studied by \cite{mateos2007} showing that only a minority (15-20\%) of the sources display spectral variability on months-years timescales. Similarly, \cite{yang2016} finds significant evidence for strongly variable column densities in only two out of 68 sources selected from the Chandra CDFS survey.

The conclusions of this study can be summarised as follows:
\begin{itemize}
\item
  All the sources investigated here show significant flux variability (with a confidence level $\geq 99\%$).
  The lowest significance of flux variability, 99\% confidence level, is found for the source with less spectral counts (source 352).
 This confirms that AGN are typically variable in the X-rays over timescales of months-years and that their variability is detectable when good enough statistics is available;
\item There is no strong evidence of spectral variability from the analysis of the hardness ratios (HR);
\item The power-law photon index variability in this sample is found less commonly than the flux variability.   We found one source with photon index variability  (source 337) which displays a 'softer-when-brighter' trend. The result is found on the basis of the HR study and on the spectral analysis. We do not find any source with a significant evidence for the opposite trend. The low fraction of spectrally variable sources is not expected to be intrinsic, since the spectral variability has been found in ensemble analyses \citep{gibson2012,serafinelli2017}. Indeed, its detection can potentially increase with the sample statistics as well as with the counts statistics.
  \item  There is no significant evidence of absorption variability across the six epochs or between the high and low states. Thus, column density changes do not explain the bulk of the flux variability observed in this study;
\item  The narrow iron line is detected in both the high and low flux states (with compatible EW) for two sources (319, 337). In three sources (203, 33, 352) the narrow line is detected only in the low state. The lower iron line EW for higher continuum luminosities can be explained as an effect of line dilution by the intrinsic continuum; 
\item The Compton reflection component is needed to fit the spectra of two sources (352 and 48) which have very high HR. We did not find any variability of this component as expected if the reflection arises in material distributed at large distances from the central engine;
\item A soft excess component is required to fit the spectra of three sources (292, 48 and 358). They are equally well fitted by a  Bremstrahlung plus line components which represents the emission from the NLR or by a scattering component (a simple power-law). We do not find any significant evidence for a variability of this component as expected if it is emitted in material located far away from the central core.
\end{itemize}

We conclude that the flux variability is a common feature of most bright/unabsorbed AGNs while the spectral variability is less strong. 
Thus, the bulk of the observed variability cannot be ascribed to changes in the primary emission mechanism nor in variations in the intervening absorbing material but to variations in the normalisation of the primary power-law continuum.

\begin{figure*}
  \centering
   \includegraphics[width=14cm]{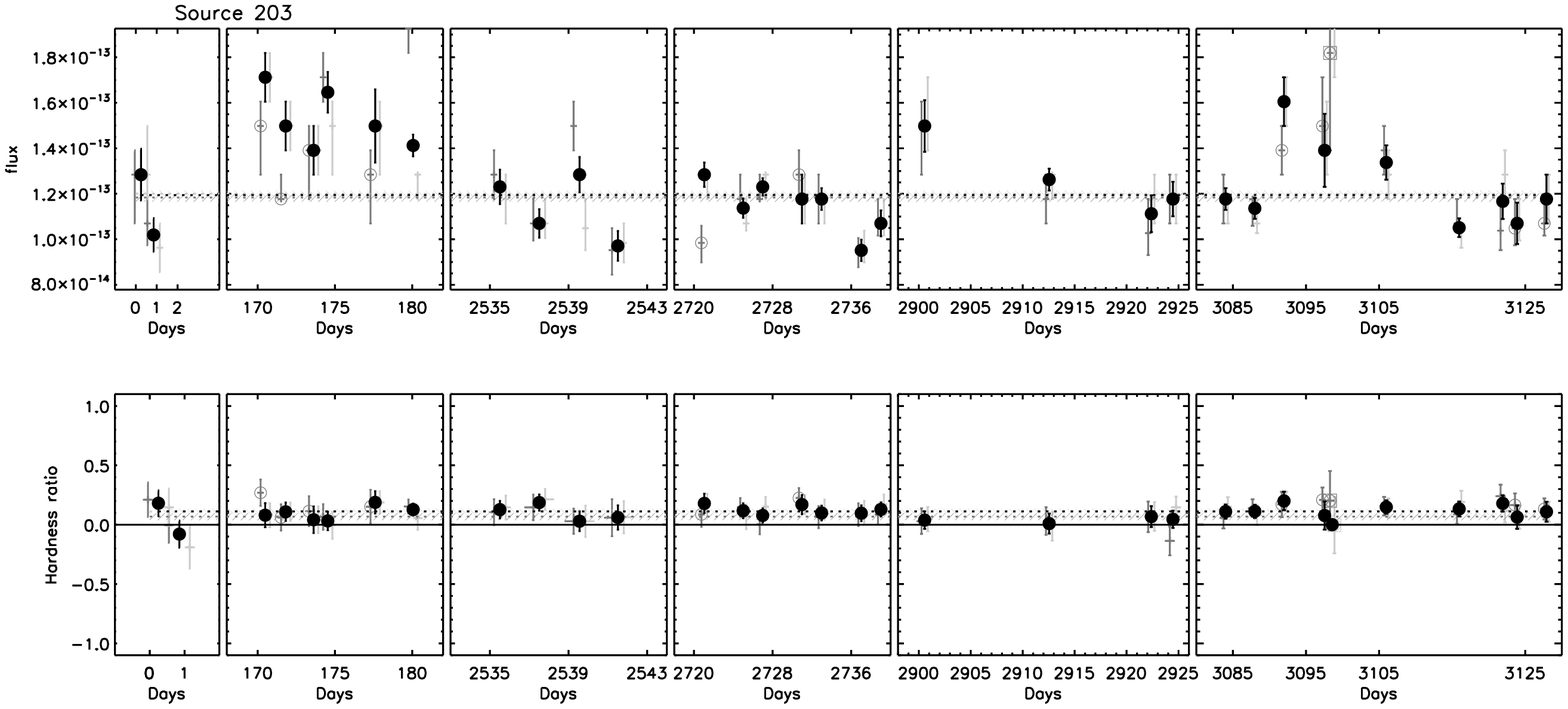}\\
   \includegraphics[width=6cm]{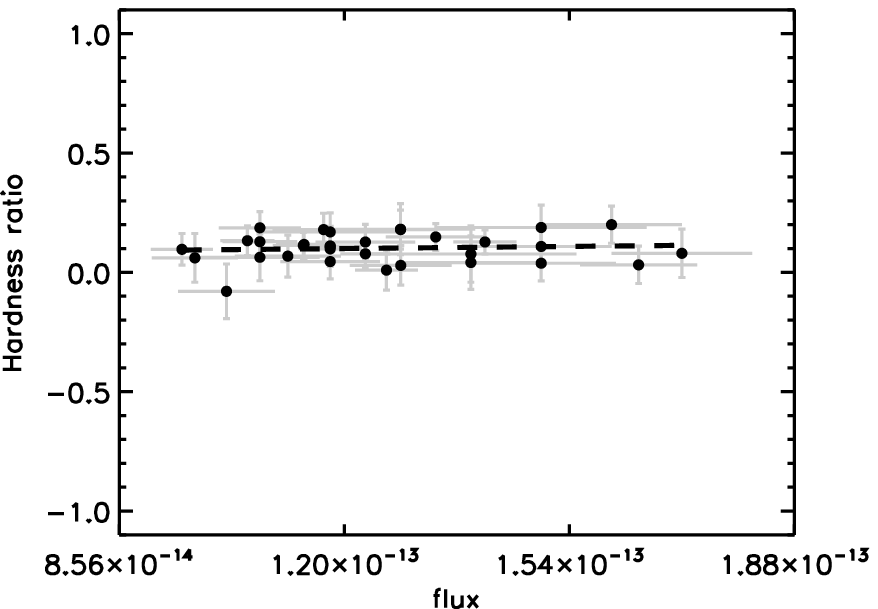}
   \includegraphics[width=6cm]{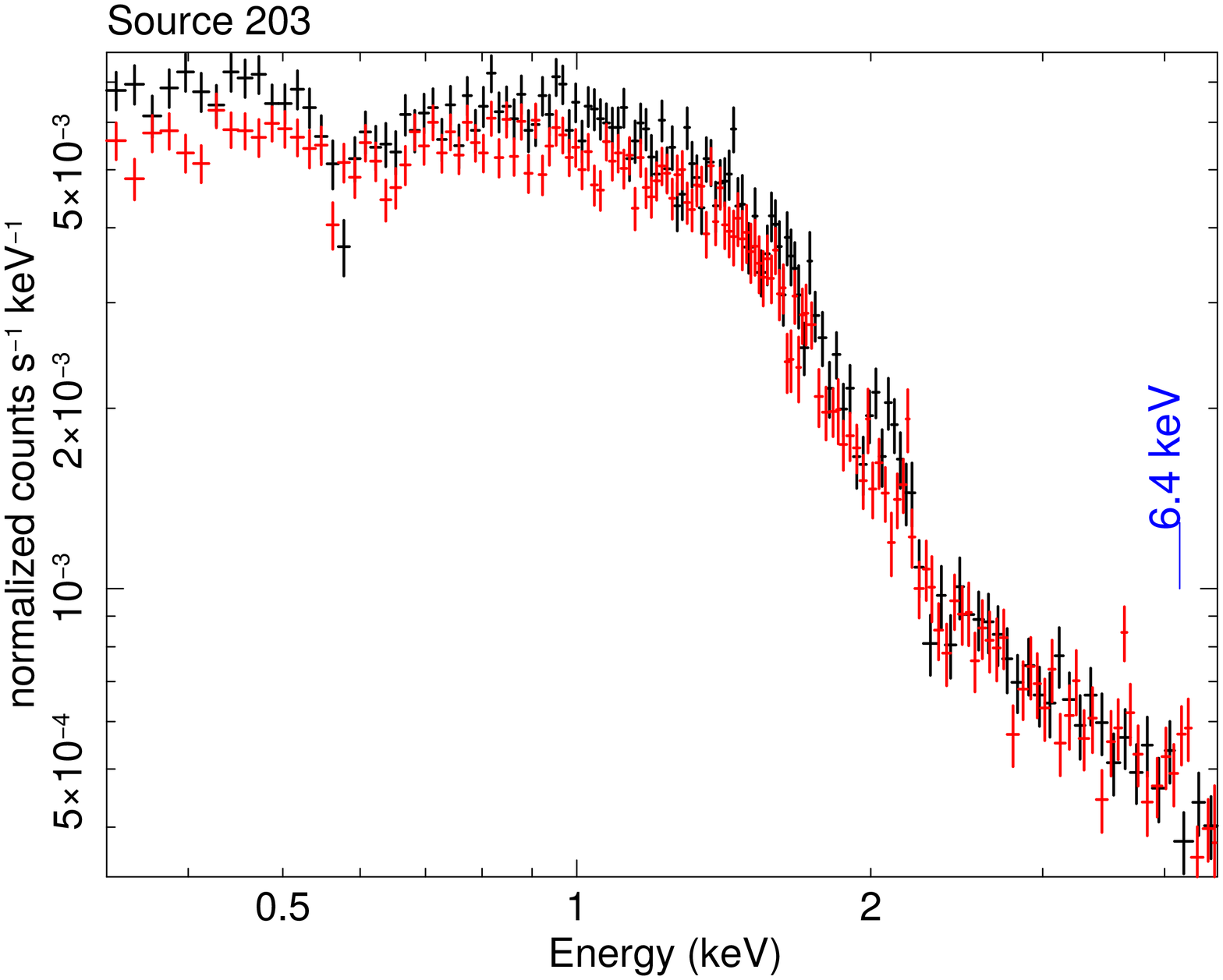}\\
    \caption{   \emph{First and second rows}: lightcurves in fluxes and HR.
     The fluxes are computed in the 0.5 and 8 keV observed frame band averaged between MOS1 and MOS2 at each of the 33 observations following the procedure of \cite{park2006}.   Black solid dots represent the weighted average of the two MOS detectors, where we only use epochs without any potential problem in the photometry.
     Observations with potential photometry problems (e.g. where either more than 10\% of the source area falls outside one of the MOS chips or on an intra-chip gap, or where the source and background regions fall on different chips) are represented by an empty circle; similarly observations where more than 30\% of the background area falls outside one of the MOS chips or on an intra-chip gap, are represented by an empty square. These epochs have been represented in grey. The horizontal
error bars represent the duration of the observation.
The flux averaged over all epochs and both MOS1 and MOS2 detectors is shown as a dotted black line.  Only epochs without any potential problem in photometry are used to compute the averages shown in the plot and used for the spectral analysis described in the text. The black solid line marks the zero level in all panels.
   The average of the MOS fluxes and hardness ratios (over the epochs without any potential problem in the photometry) is represented as a dotted black line and its error is represented by dashed grey lines. 
\emph{3rd row, left}: HR versus flux of the individual observations. The dashed line is the linear regression line (calculated not considering the errors). 
\emph{3rd row, right}: high flux state and low flux state spectra represented with black and red data points respectively. }
         \label{Fig203fluxes}
   \end{figure*}

 \begin{figure*}
\centering
   \includegraphics[width=14cm]{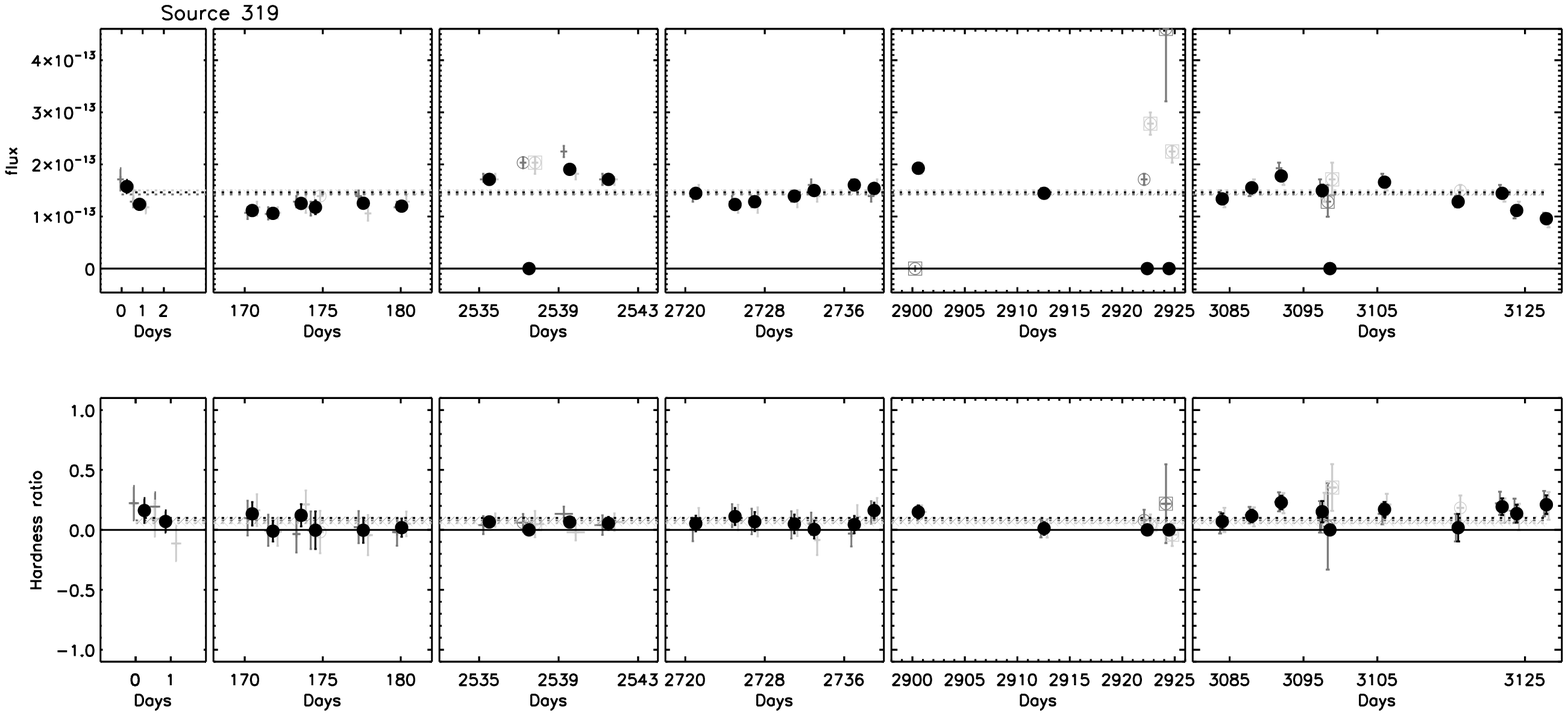}\\
      \includegraphics[width=6cm]{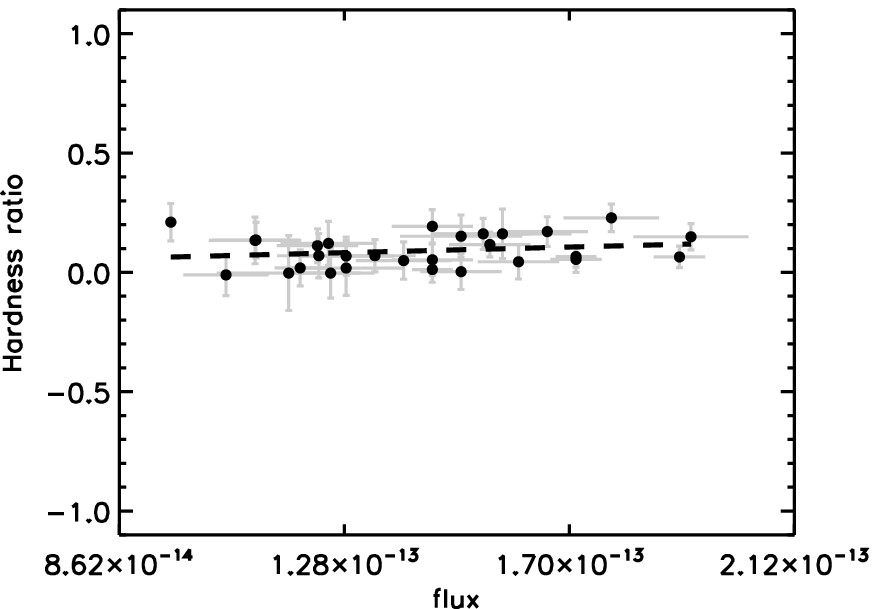}
   \includegraphics[width=6cm]{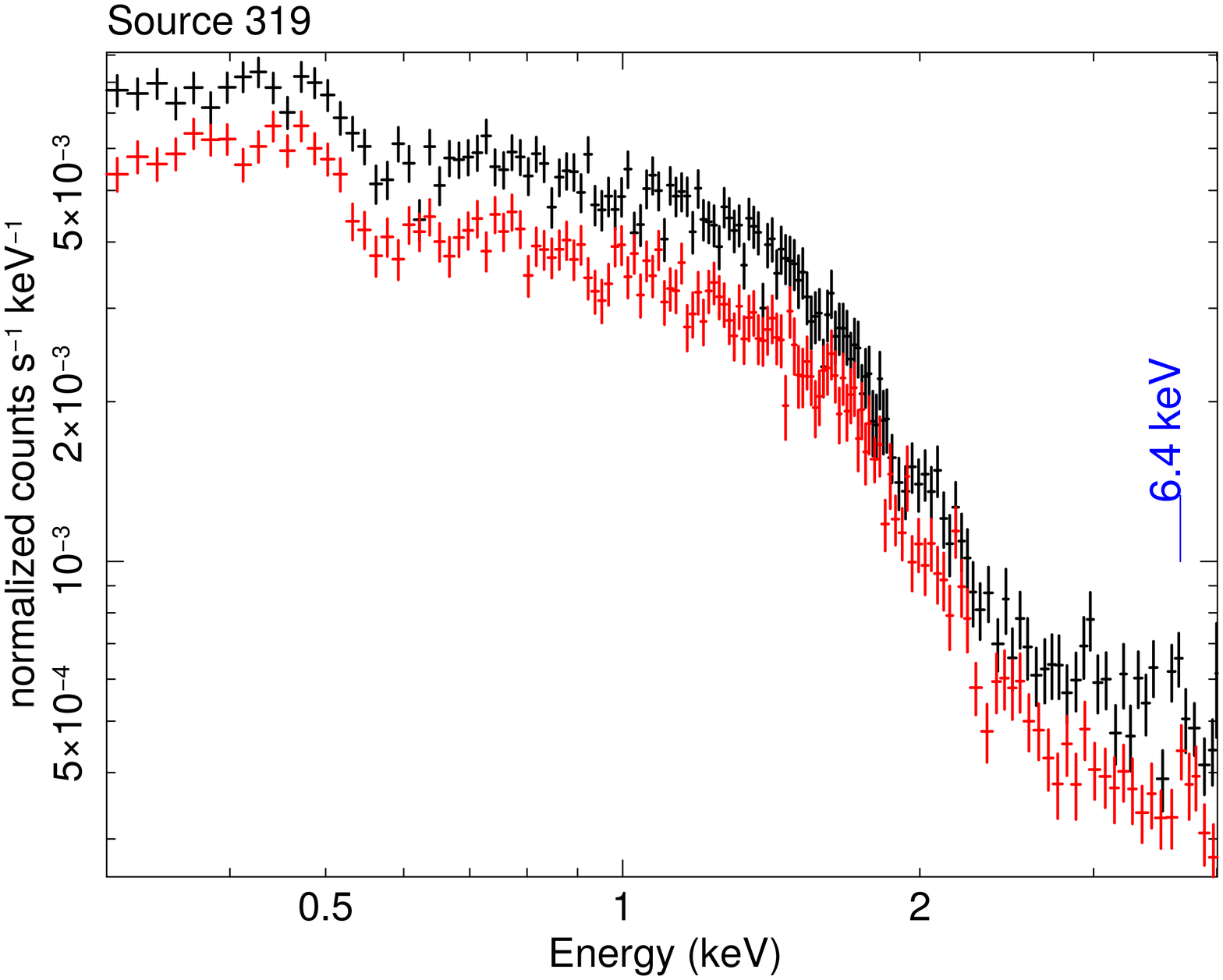}
    \caption{Source 319. Plot panels as in Fig. 2.   }
         \label{Fig319fluxes}
   \end{figure*}

 \begin{figure*}
  \centering
   \includegraphics[width=14cm]{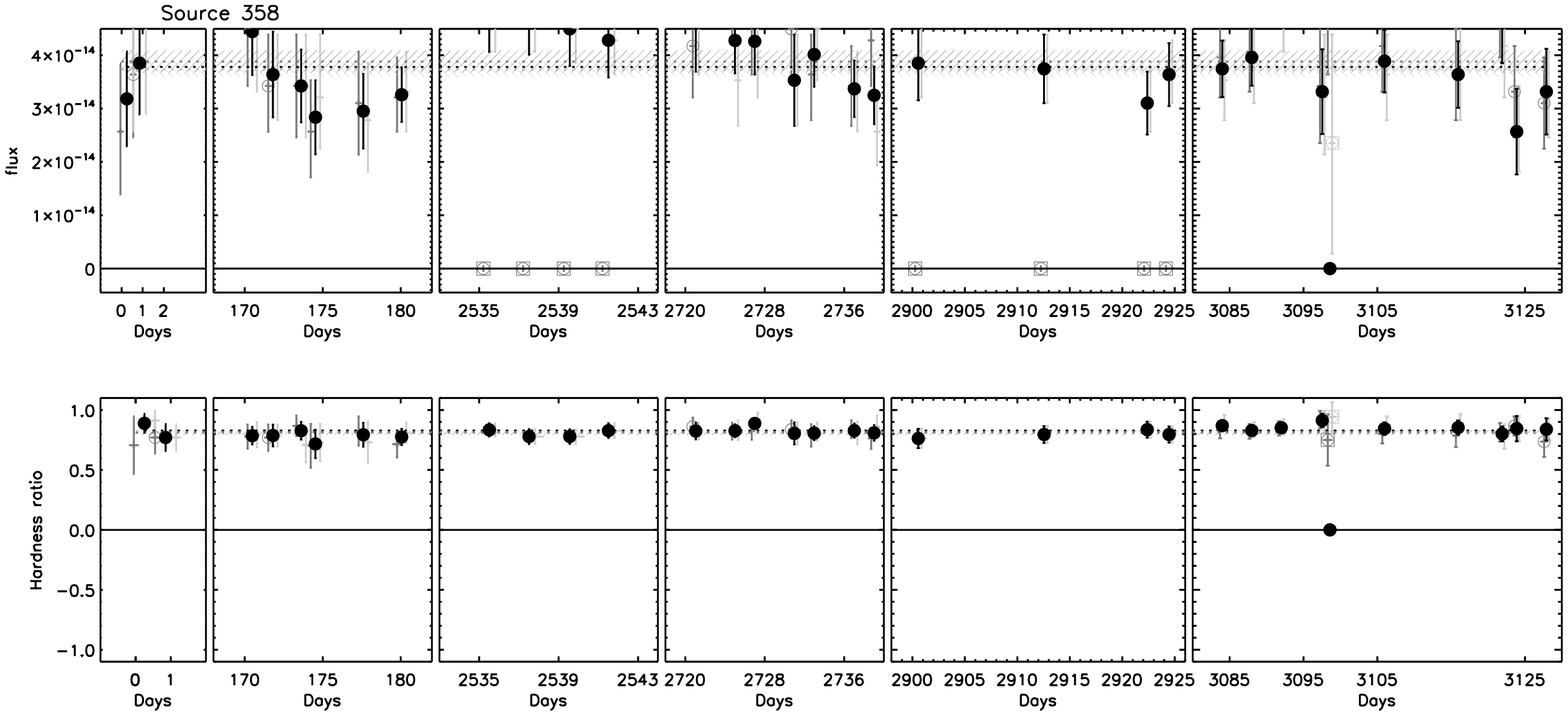}\\
      \includegraphics[width=6cm]{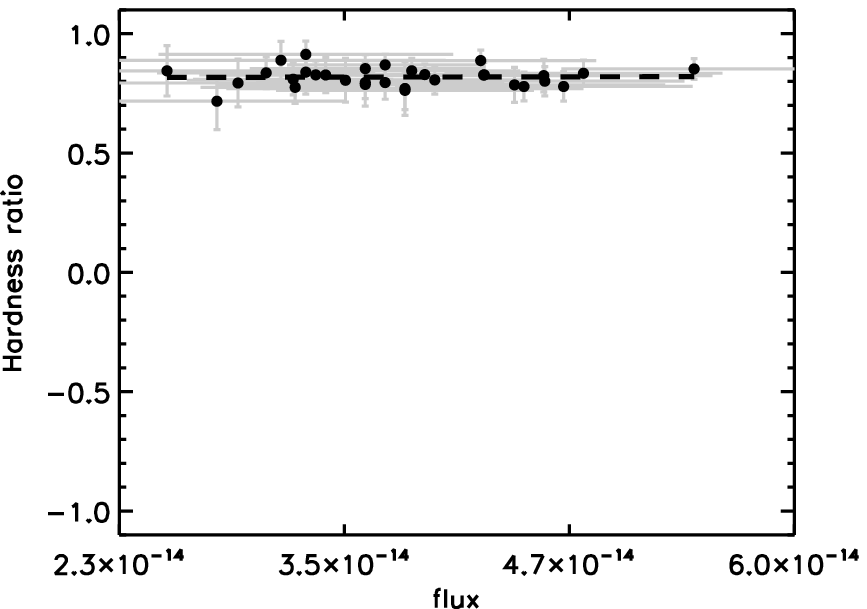} 
   \includegraphics[width=6cm]{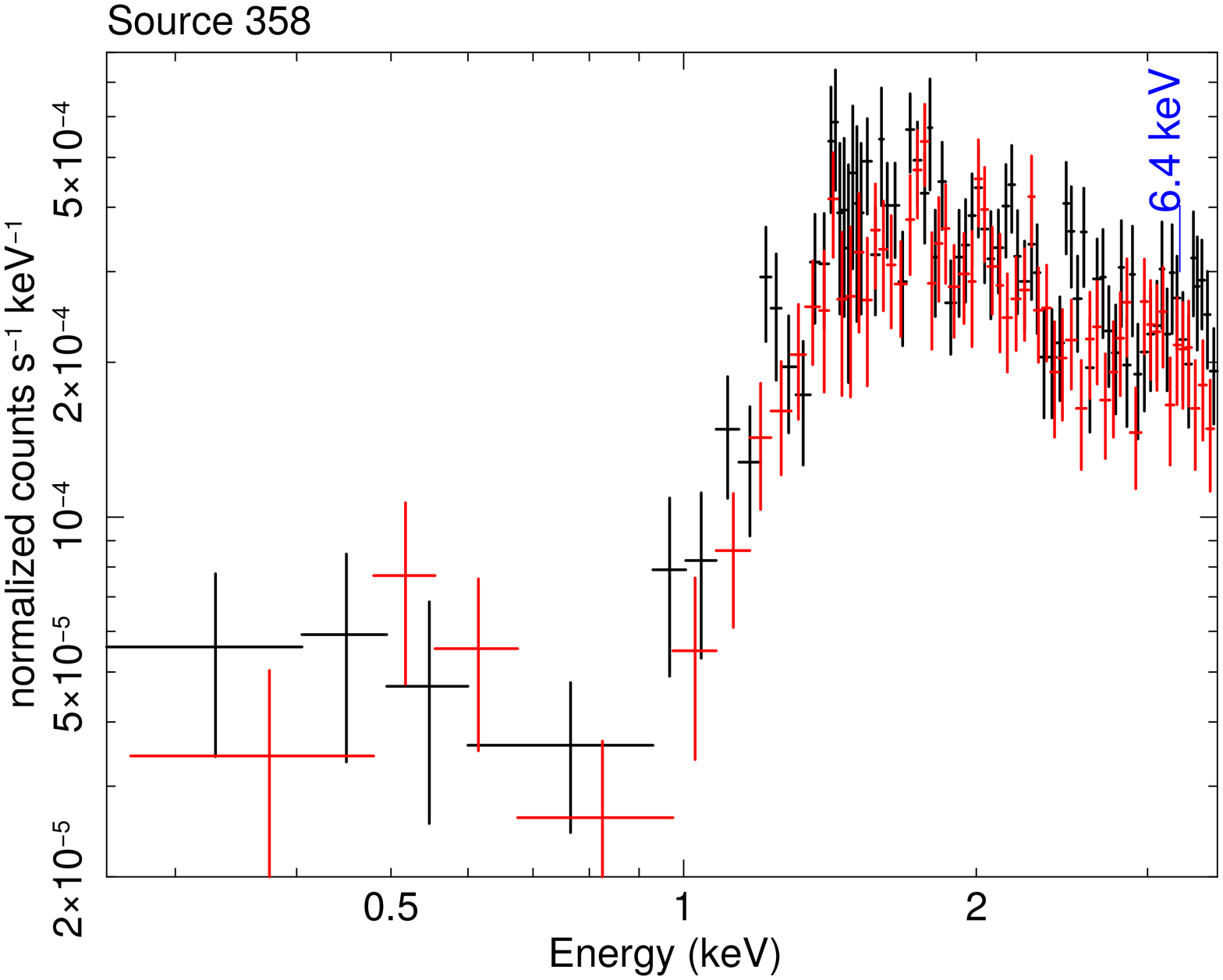}  \\
    \caption{Source 358.
      Plots as in Fig. 2.
    }
         \label{Fig358fluxes}
   \end{figure*}

 \begin{figure*}
\centering
   \includegraphics[width=14cm]{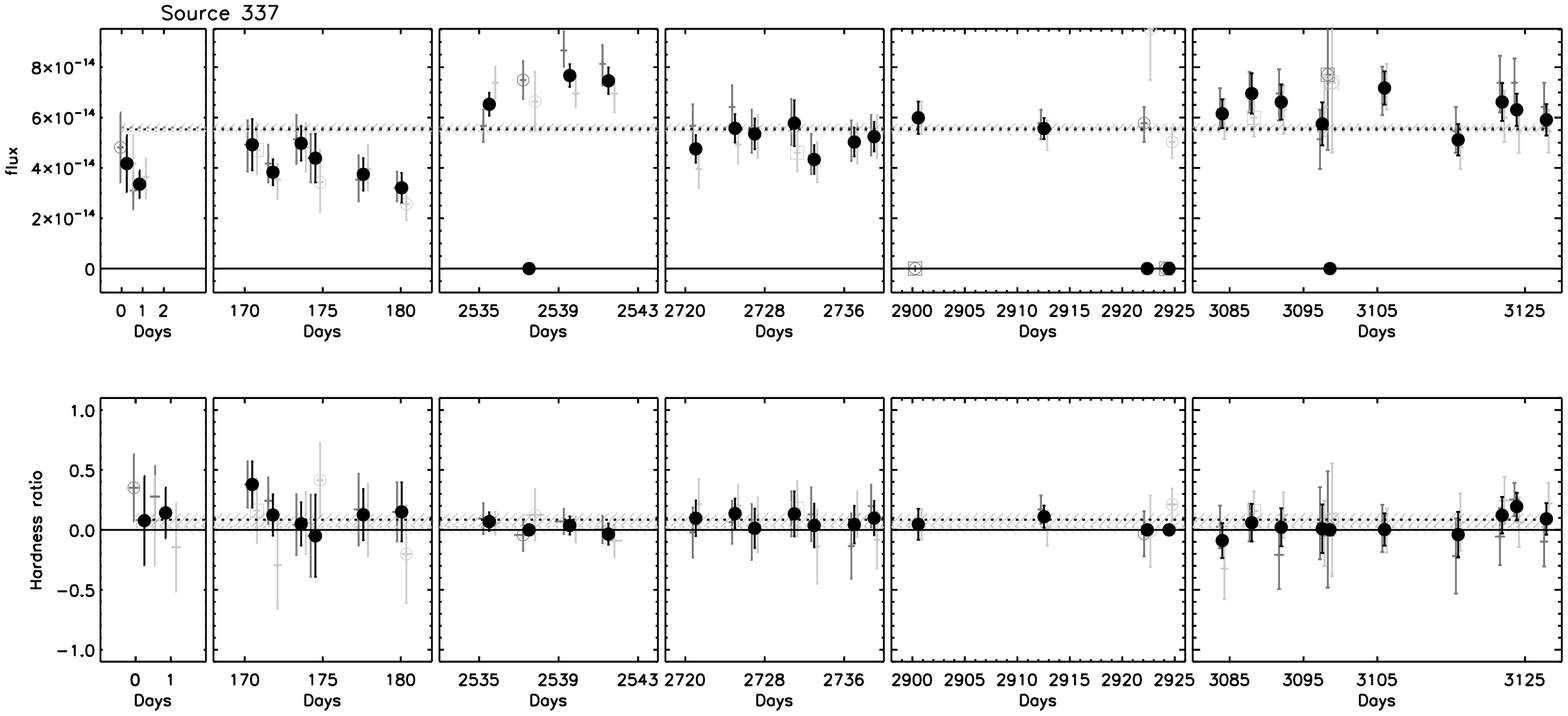}\\
     \includegraphics[width=6cm]{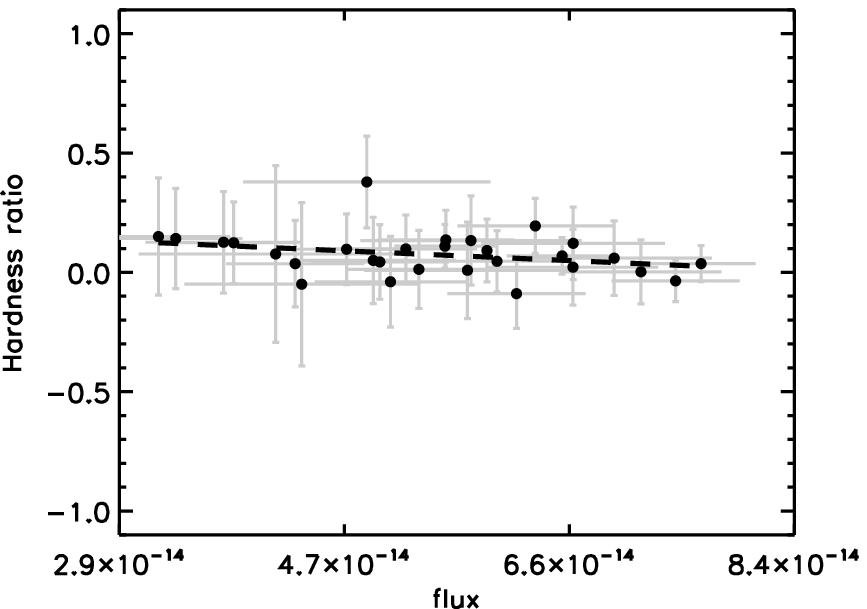}\\
  \includegraphics[width=6cm]{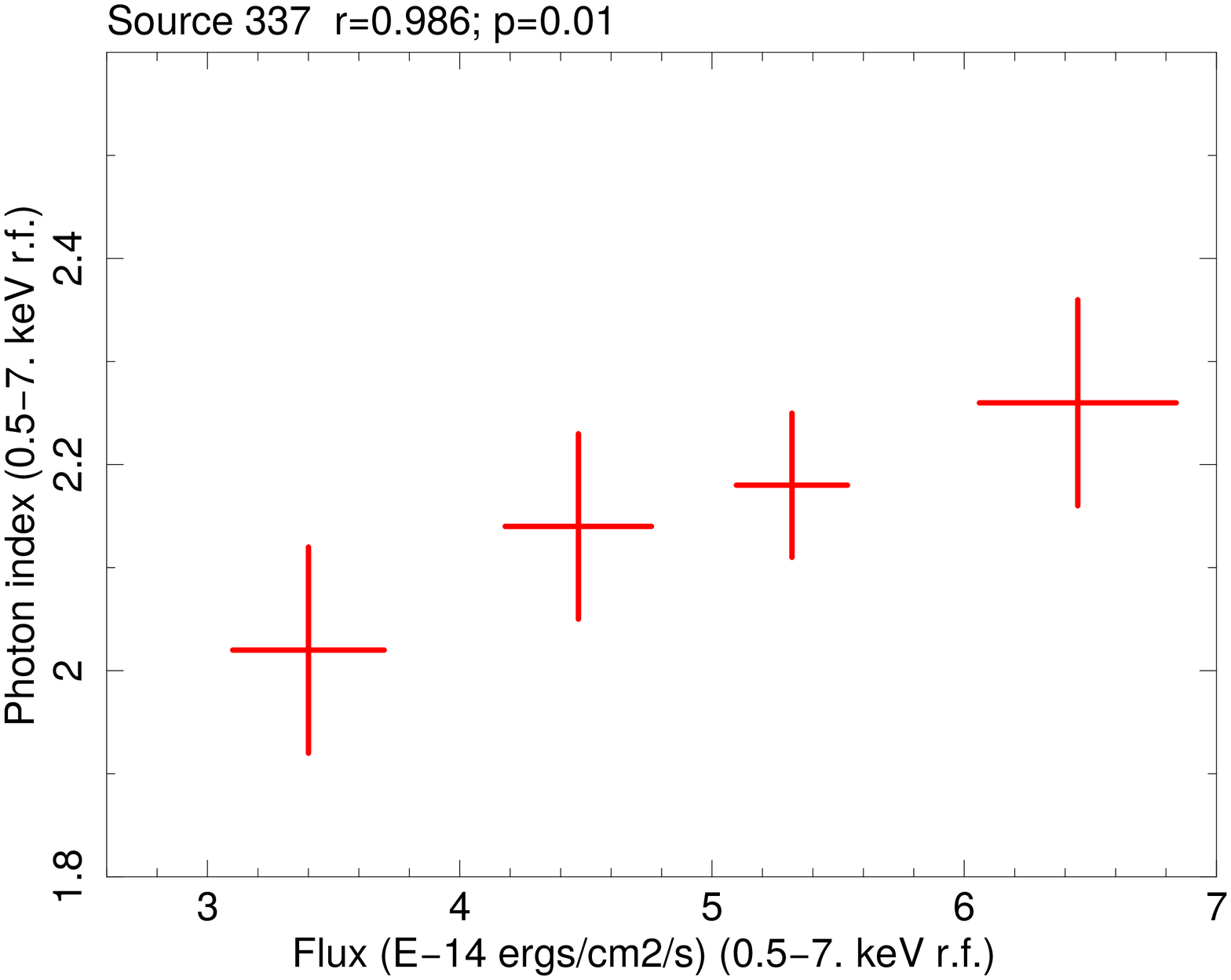}
  \includegraphics[width=6cm]{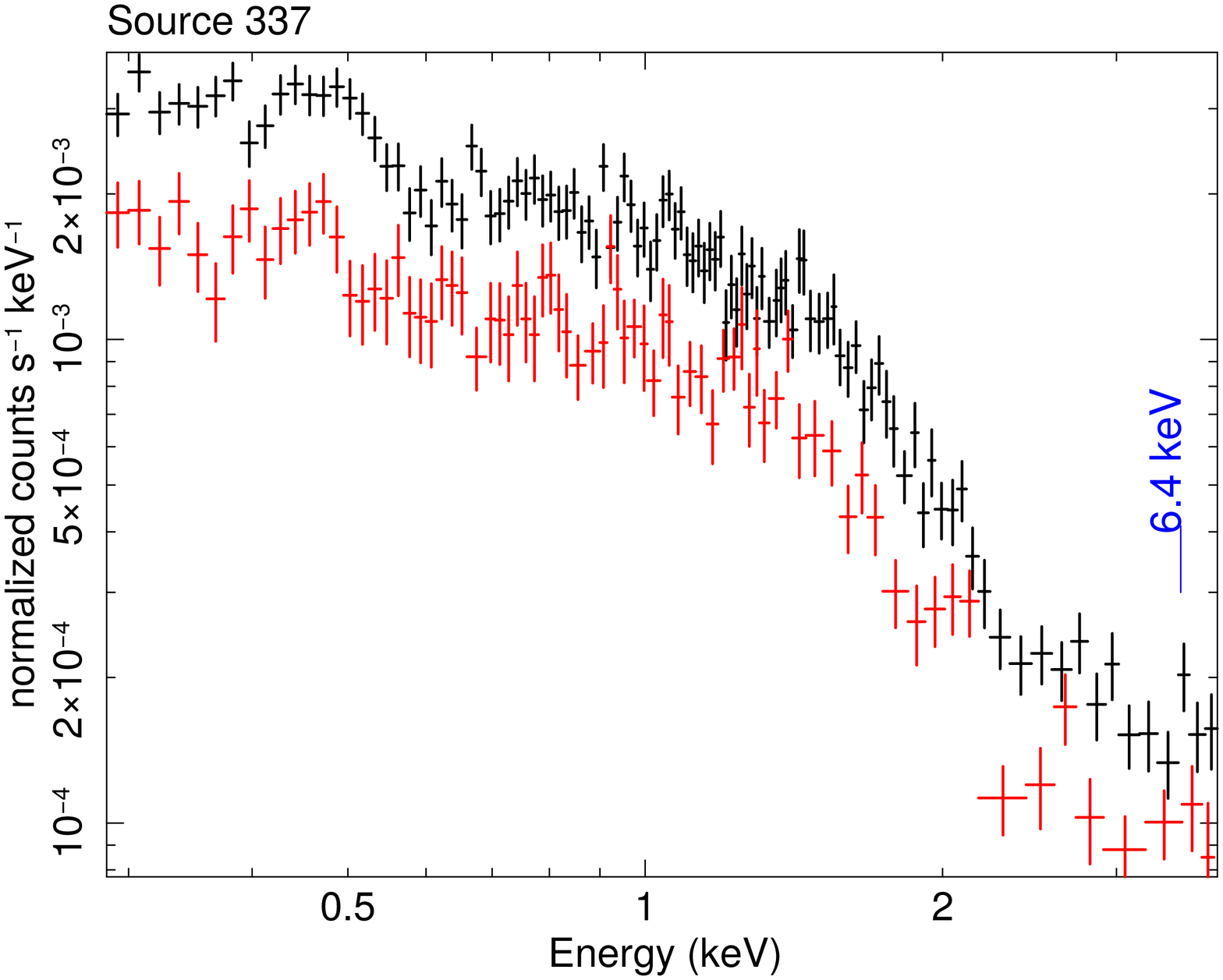}\\
\caption{Source 337. Panels from 1st to 3rd displayed as in Fig. 2. \emph{4th row, left}: powerlaw photon index versus continuum flux from the individual epochs (the first and 5th epoch have not been represented because of their low statistics).
              }
         \label{Fig337fluxes}
   \end{figure*}

\begin{figure*}
 \centering
  \includegraphics[width=14cm]{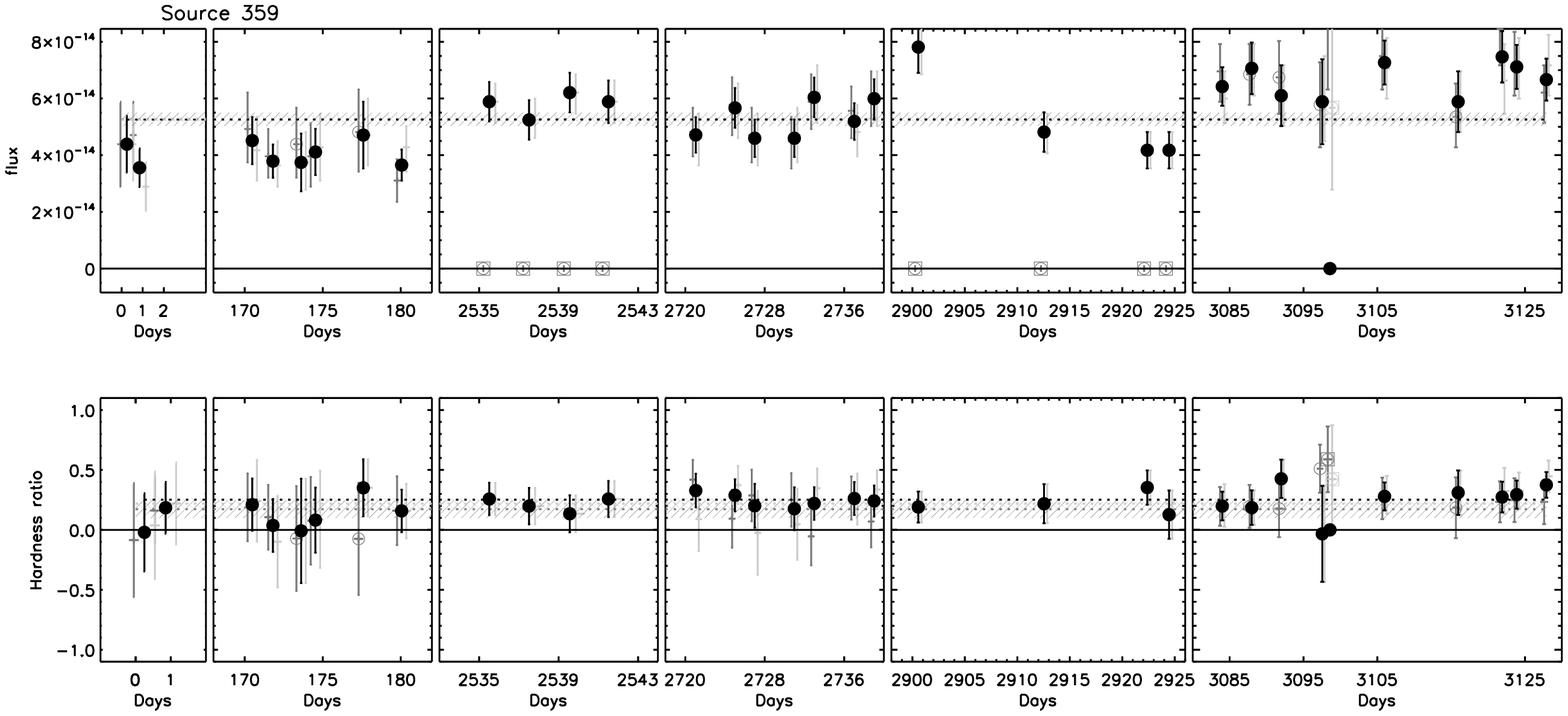}\\
    \includegraphics[width=6cm]{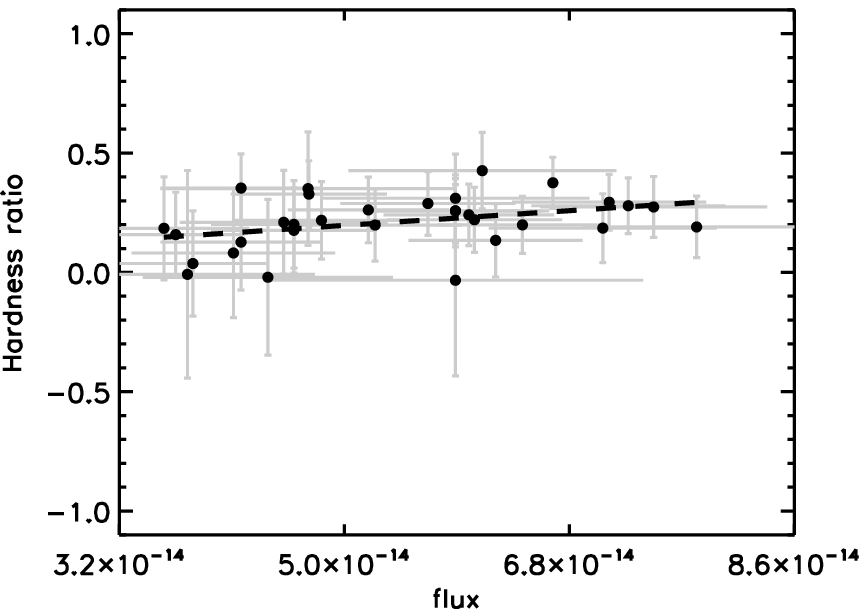}
   \includegraphics[width=6cm]{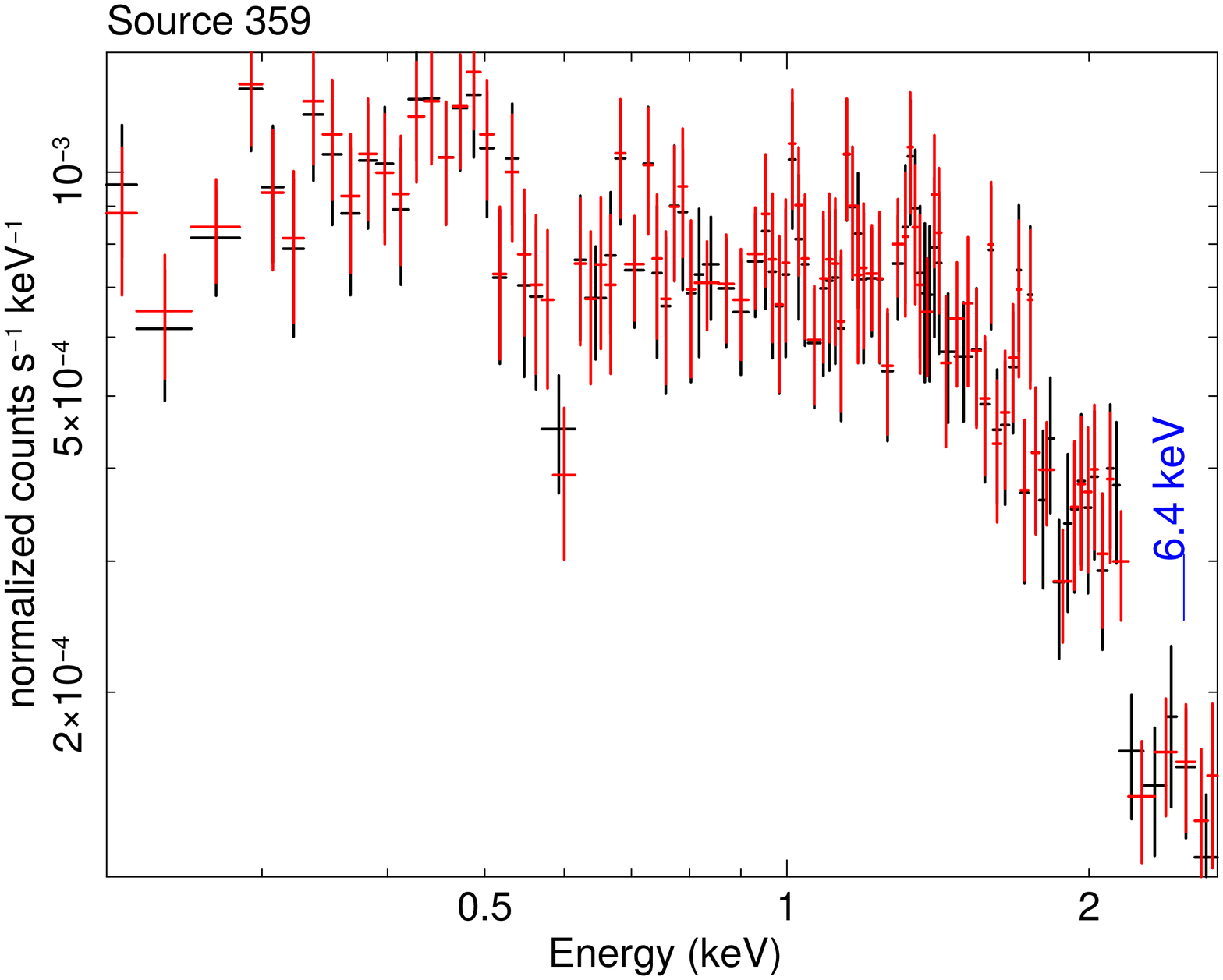}
    \caption{Source 359. Panels as in Fig. 2. 
              }
         \label{Fig359fluxes}
   \end{figure*}

 \begin{figure*}
     \centering
   \includegraphics[width=14cm]{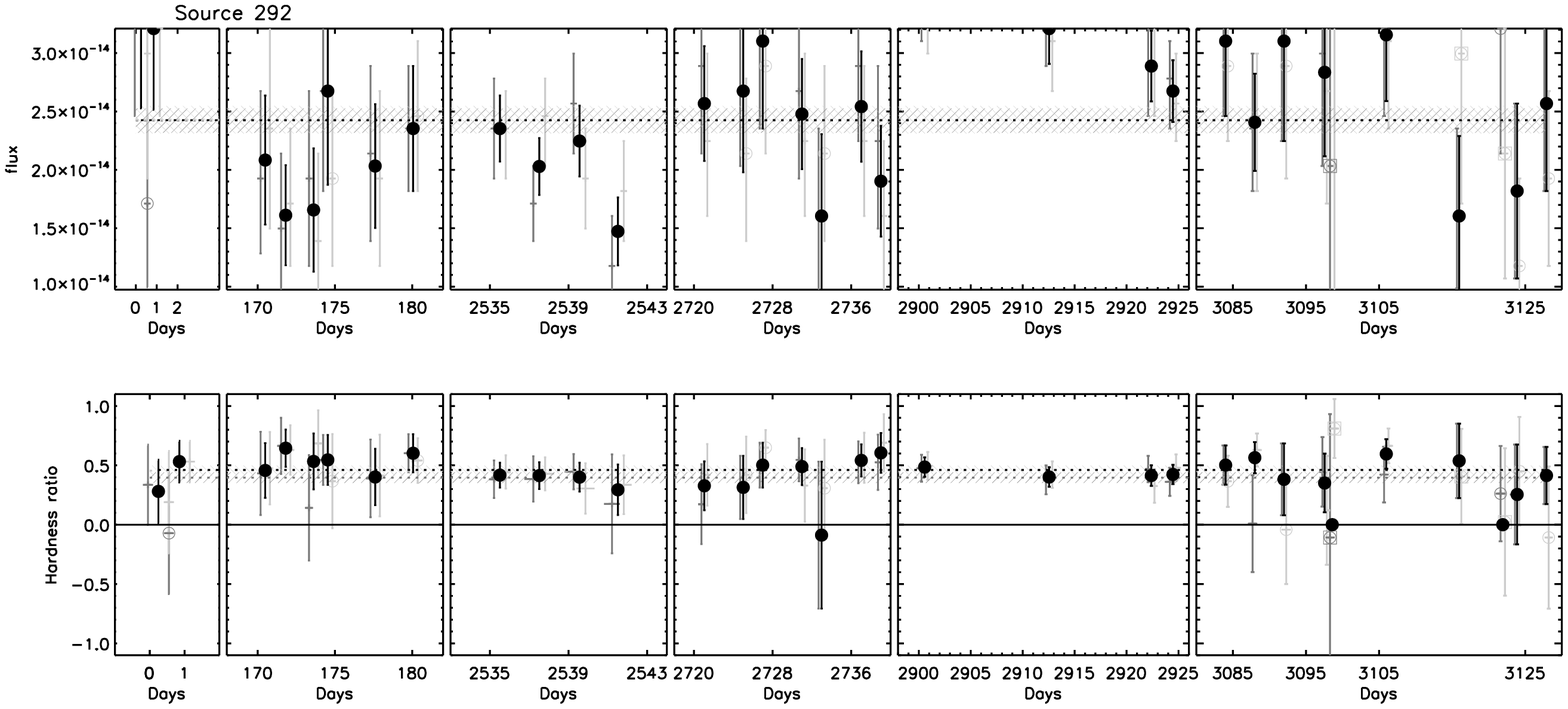}  \\
      \includegraphics[width=6cm]{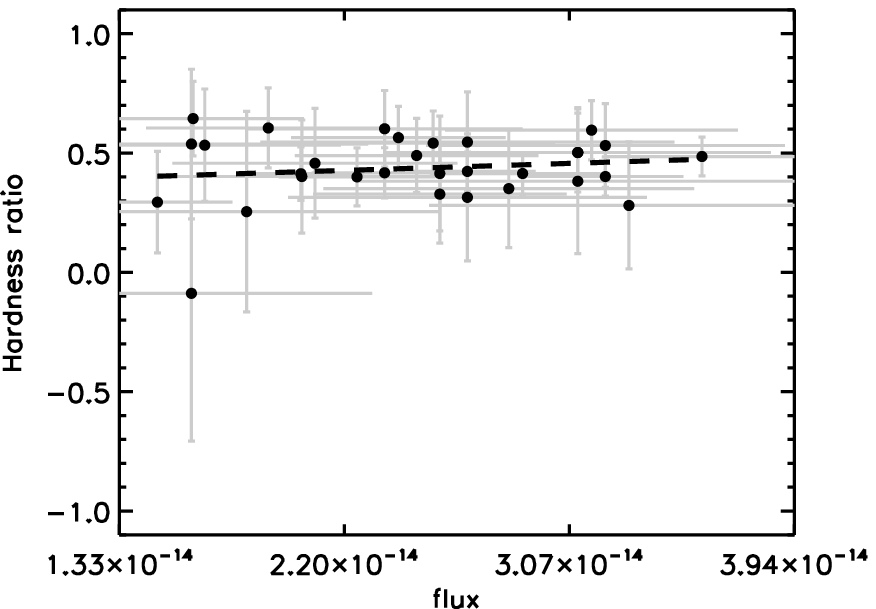}  
    \includegraphics[width=6cm]{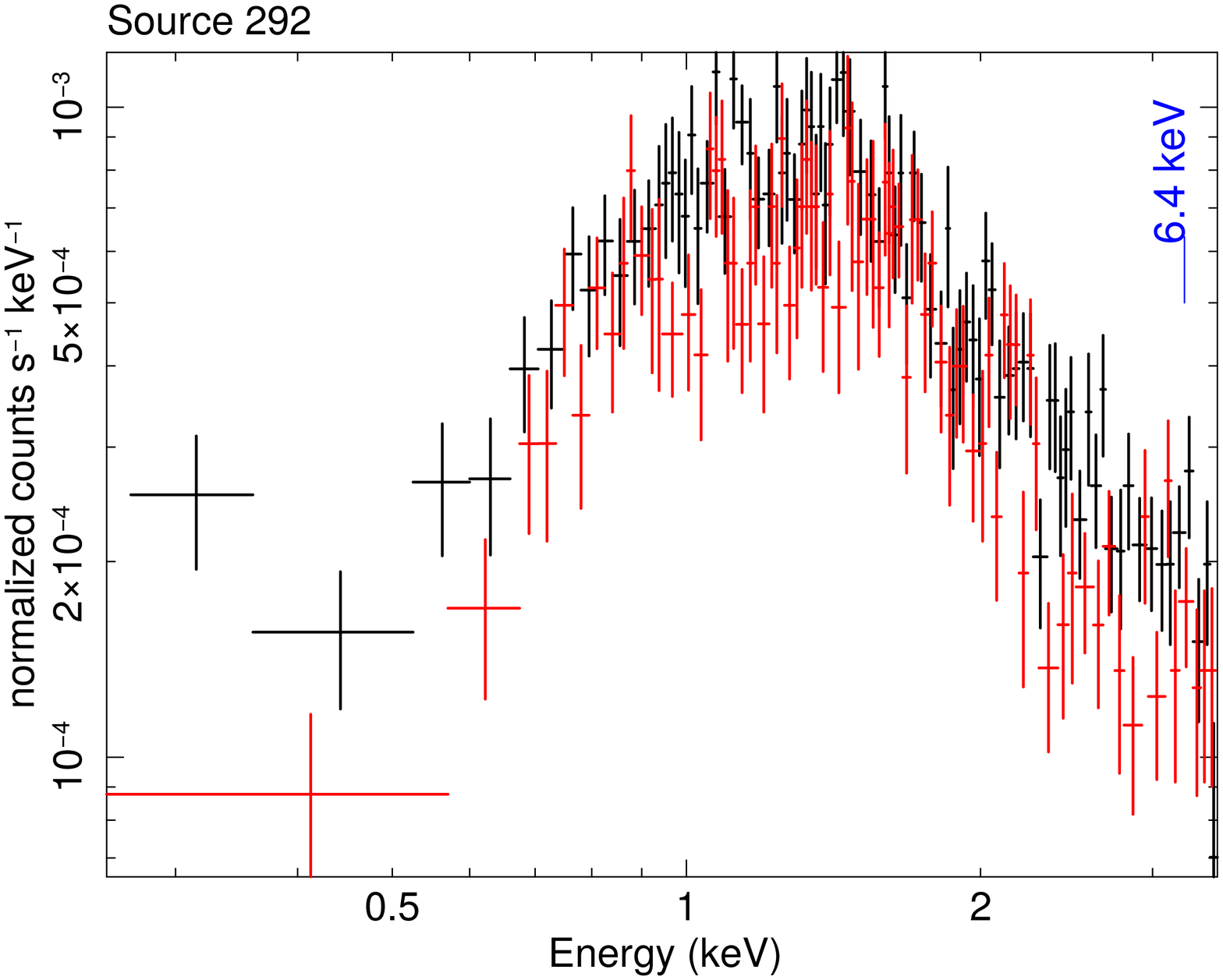} \\
    \caption{Source 292. 
      The panels are displayed as in Fig. 2.
    }
         \label{Fig292fluxes}
   \end{figure*}

 \begin{figure*}
\centering
   \includegraphics[width=14cm]{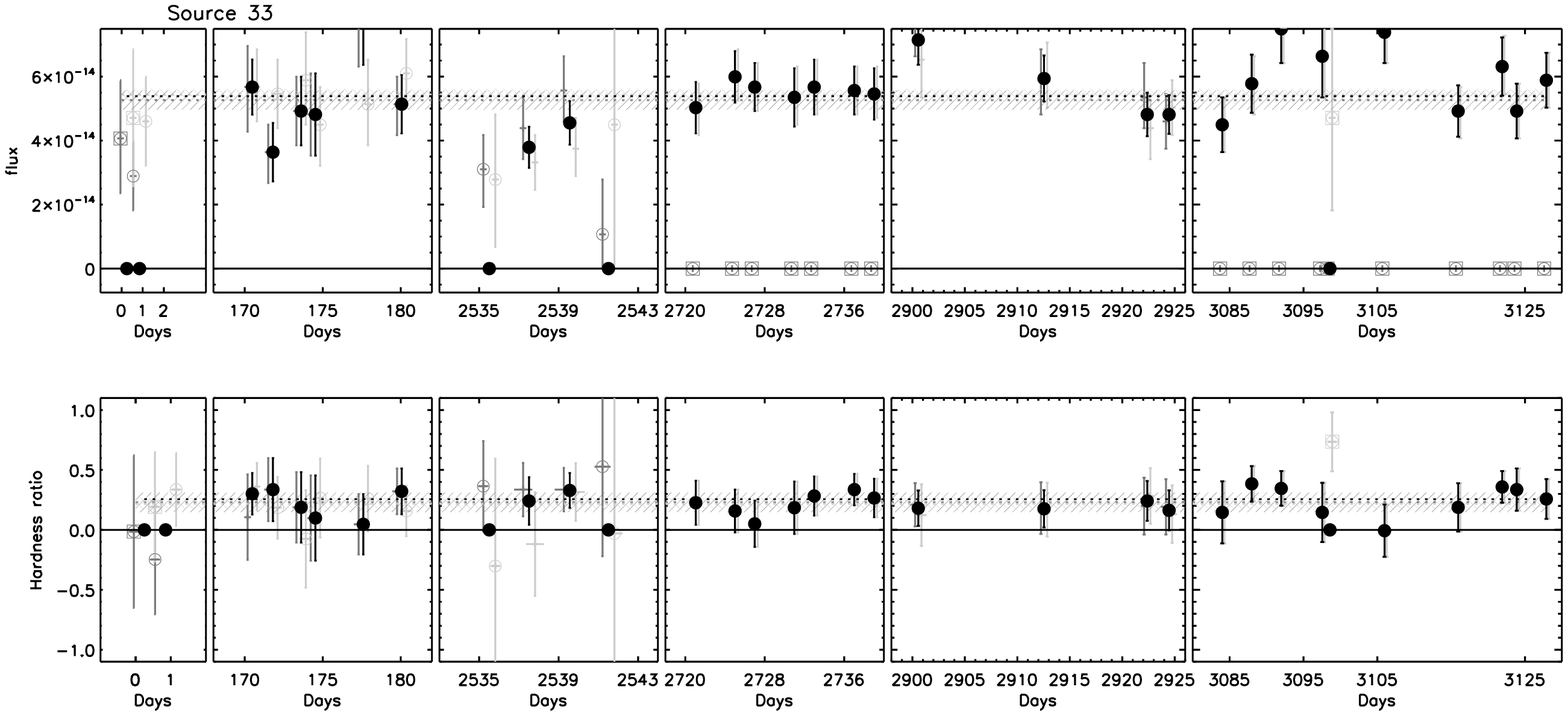}\\
      \includegraphics[width=6cm]{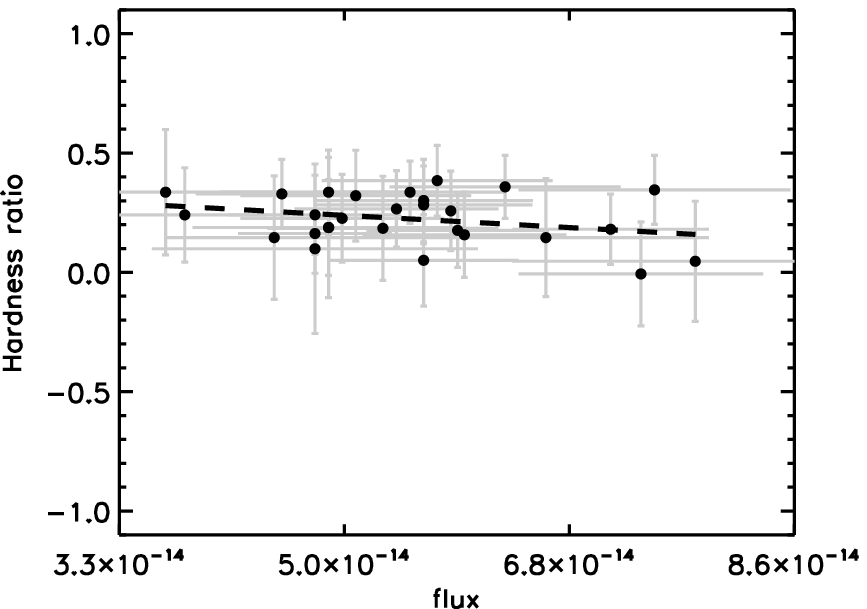}
   \includegraphics[width=6cm]{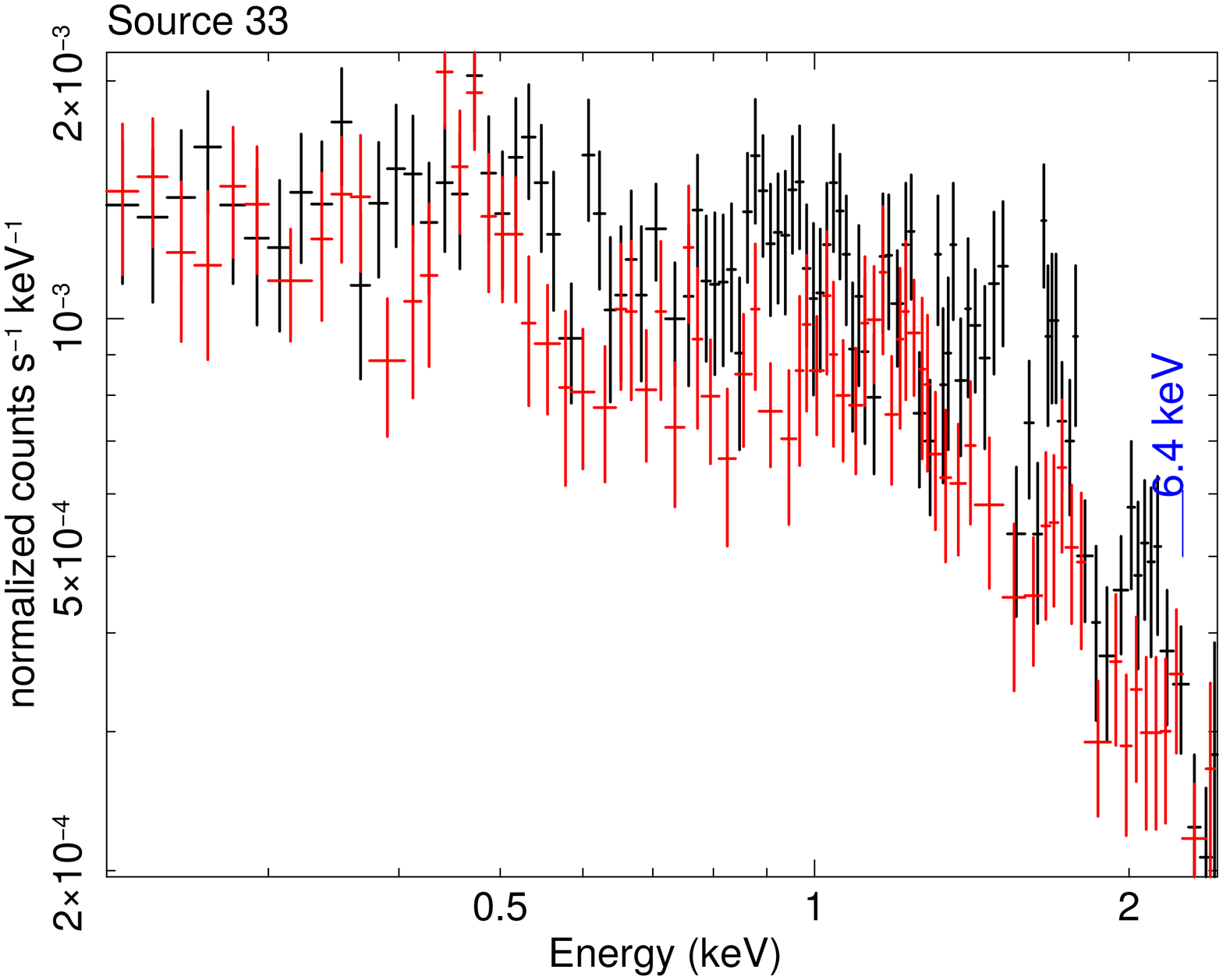}\\
   \caption{Source 33.  Panels displayed as in Fig. 2.
}
         \label{Fig33fluxes}
   \end{figure*}

 \begin{figure*}
\centering
   \includegraphics[width=11cm]{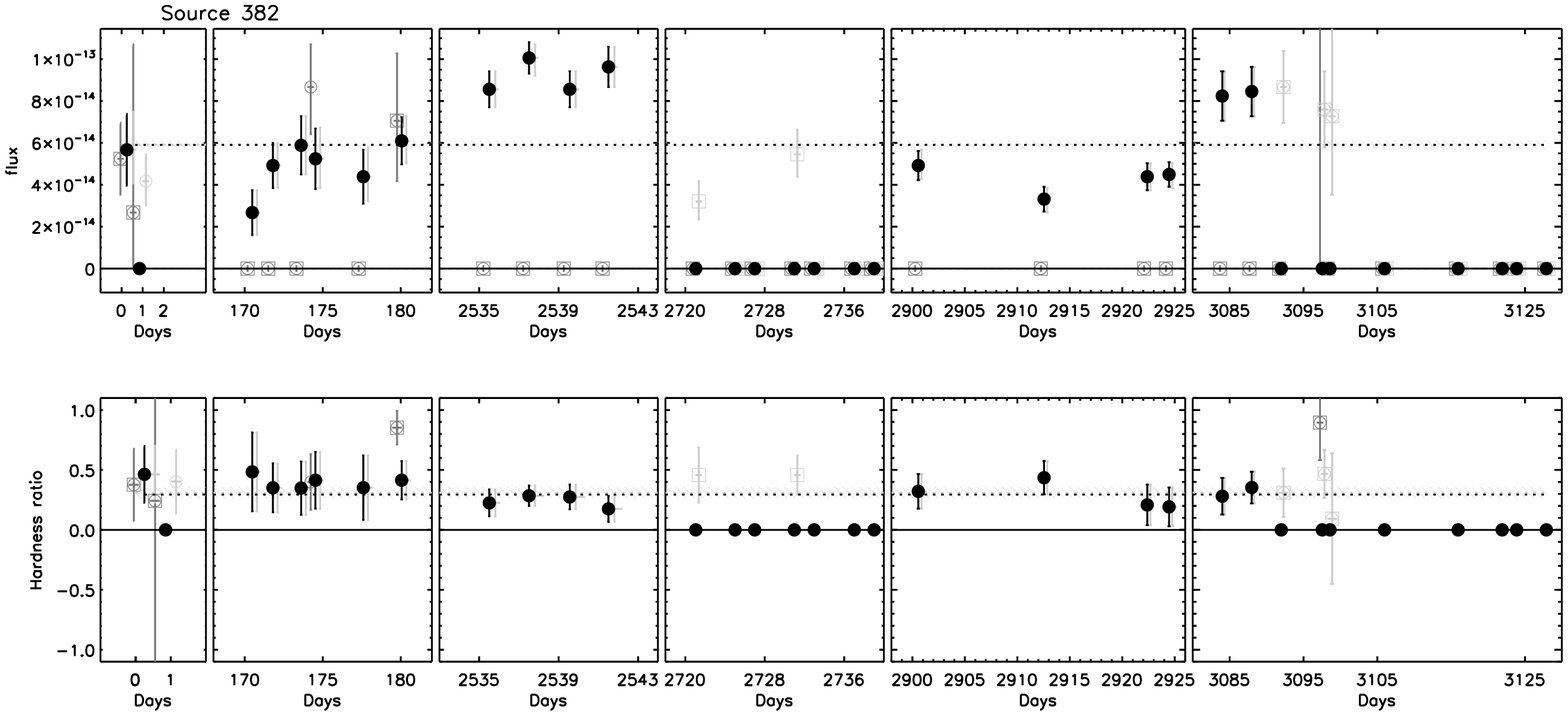} \\
      \includegraphics[width=6cm]{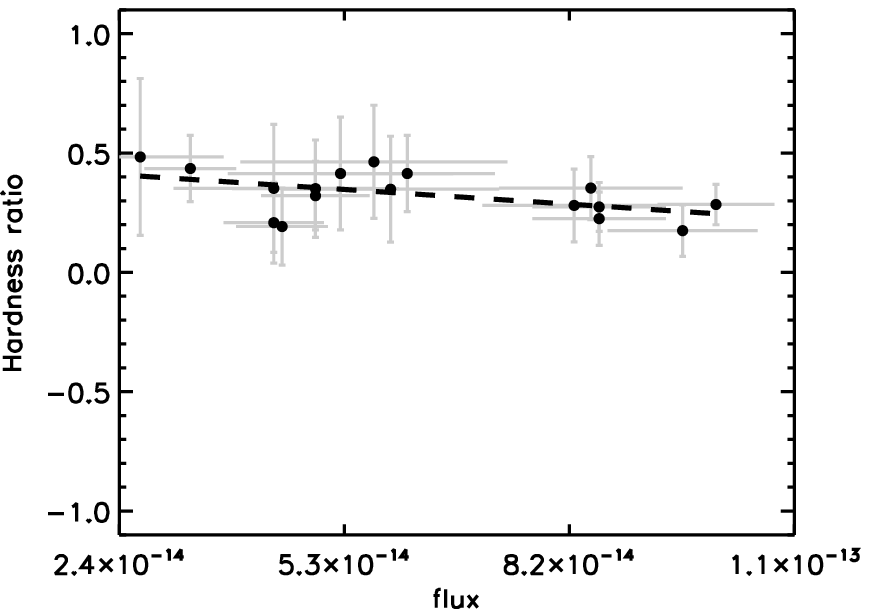} 
   \includegraphics[width=6cm]{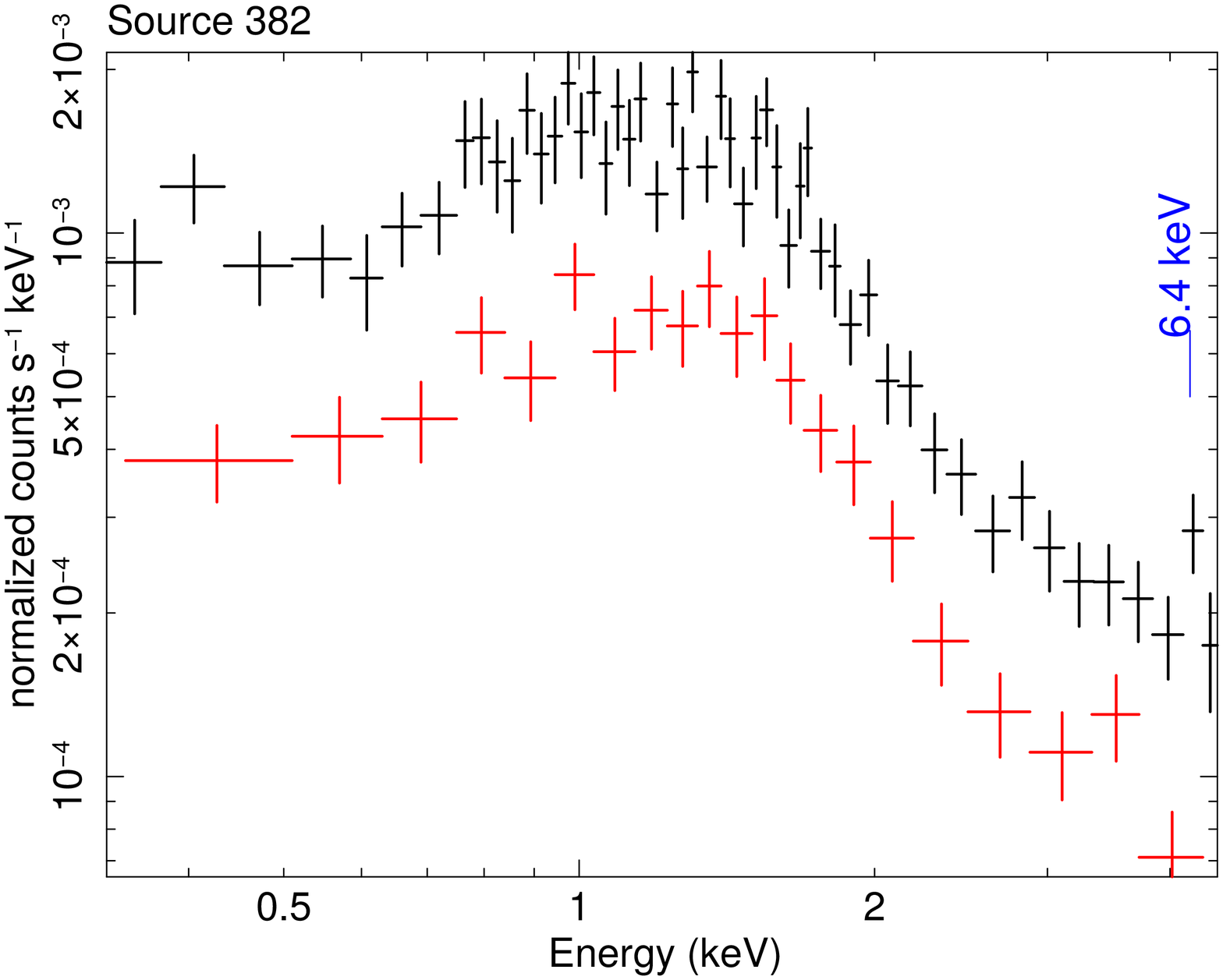} \\
\caption{Source 382. Panels as in Fig. 2. }
         \label{Fig382fluxes}
   \end{figure*}

 \begin{figure*}
  \centering
   \includegraphics[width=14cm]{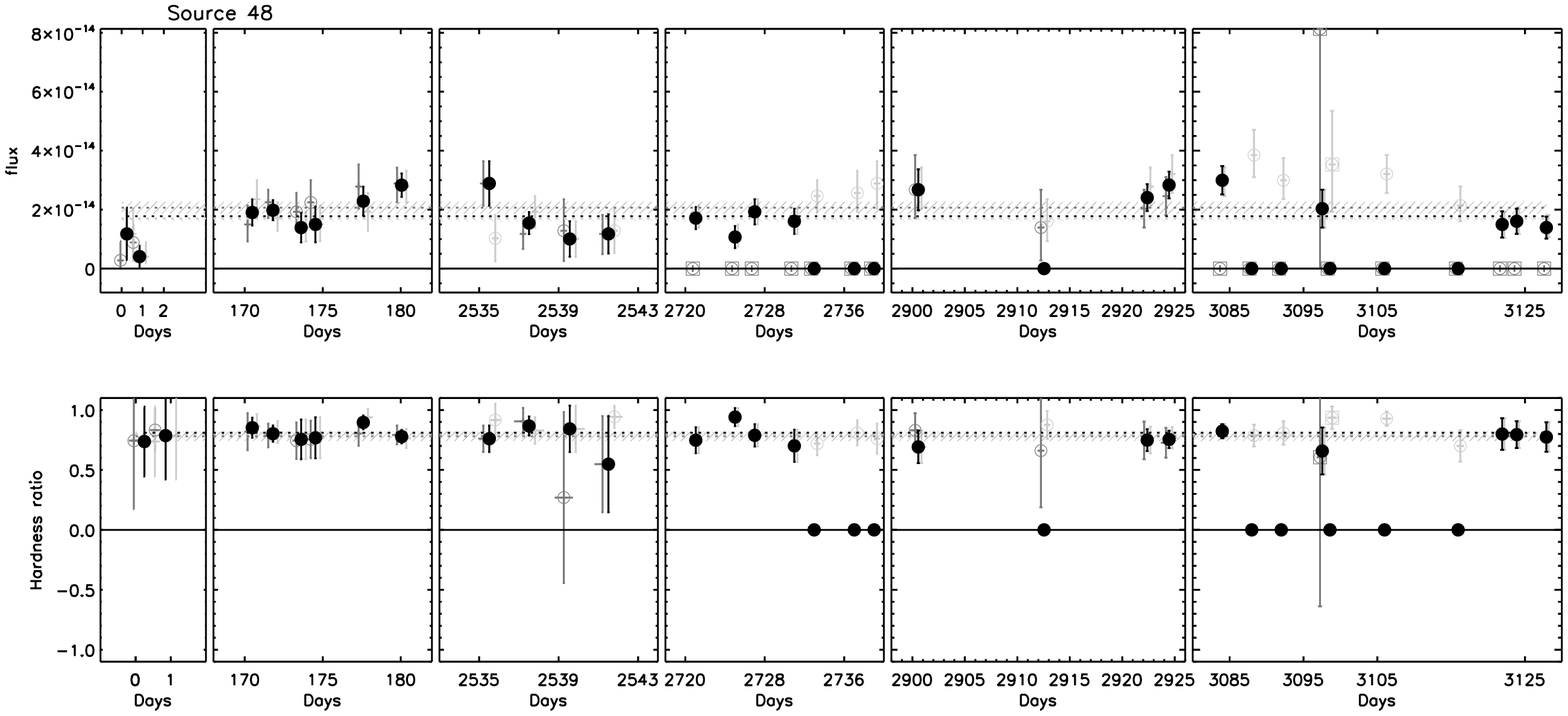}\\
      \includegraphics[width=6cm]{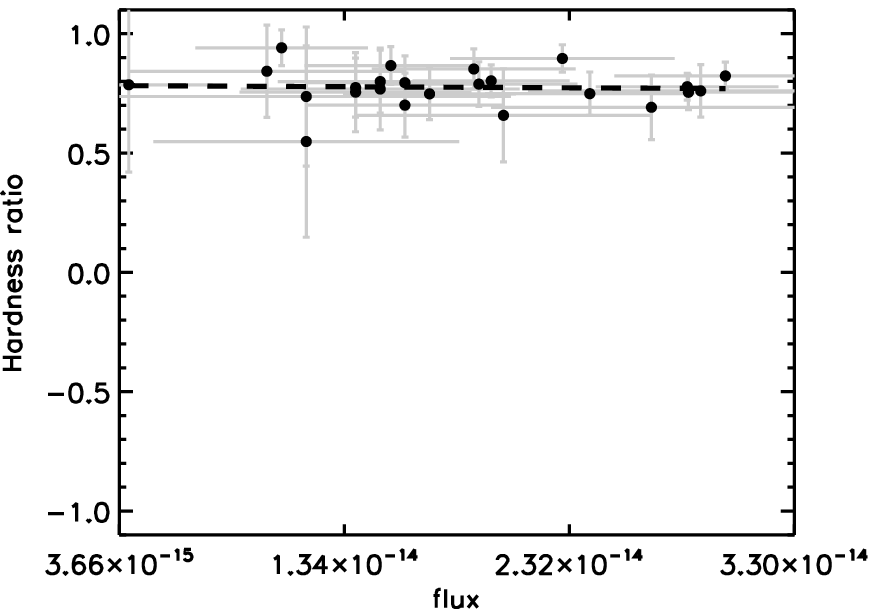}
     \includegraphics[width=6cm]{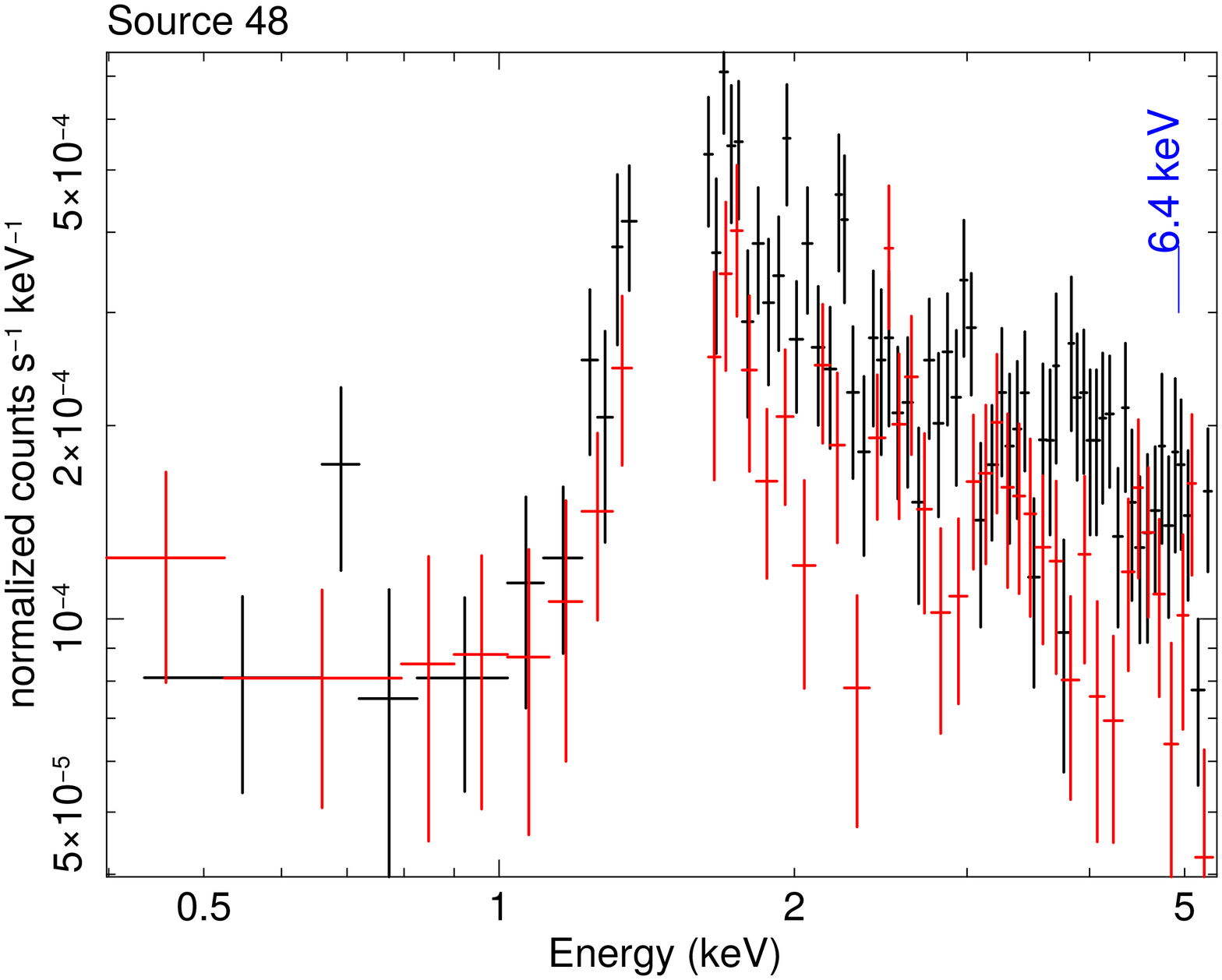} \\
  \caption{Source 48.  Plots as in previous figures. 
              }
         \label{Fig48fluxes}
   \end{figure*}

 \begin{figure*}
   \centering
   \includegraphics[width=14cm]{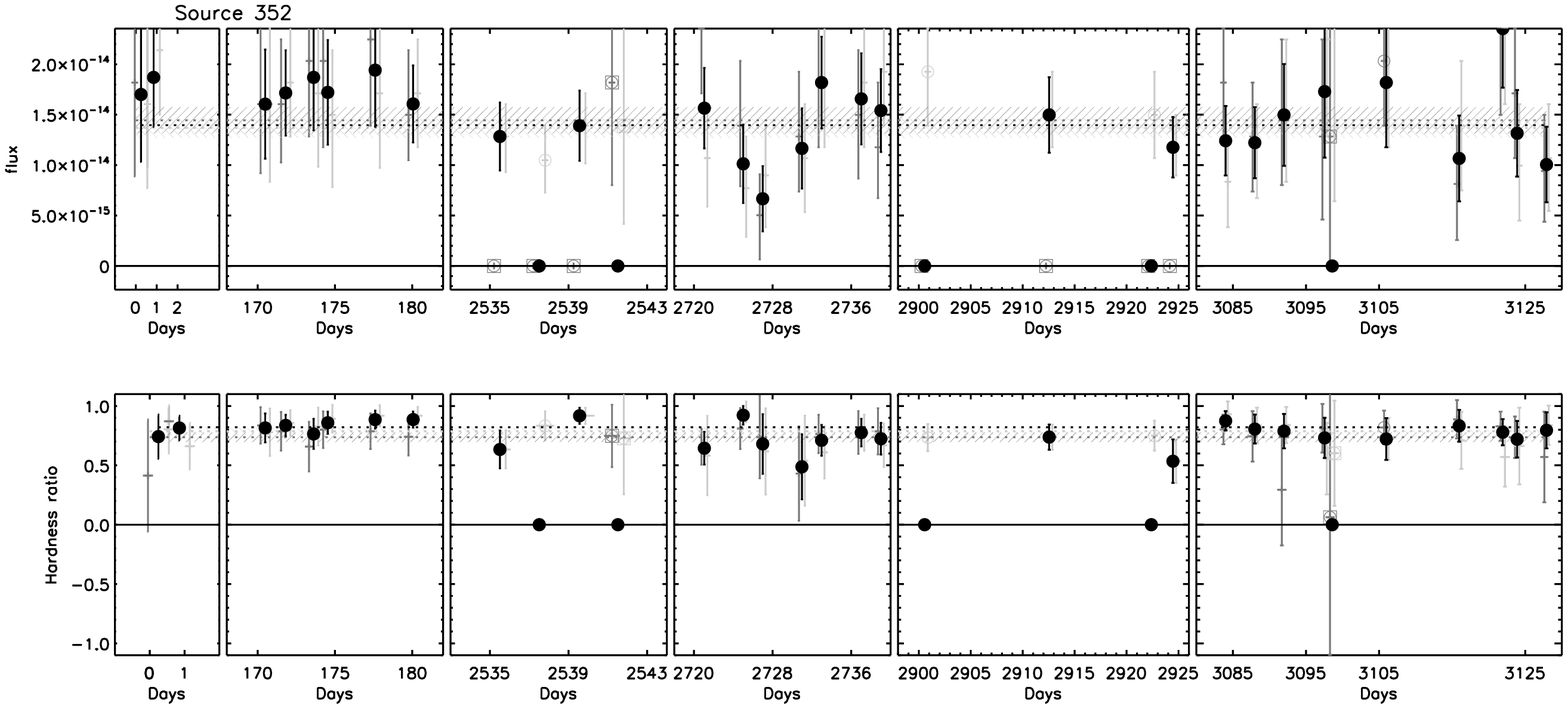}\\
   \includegraphics[width=6cm]{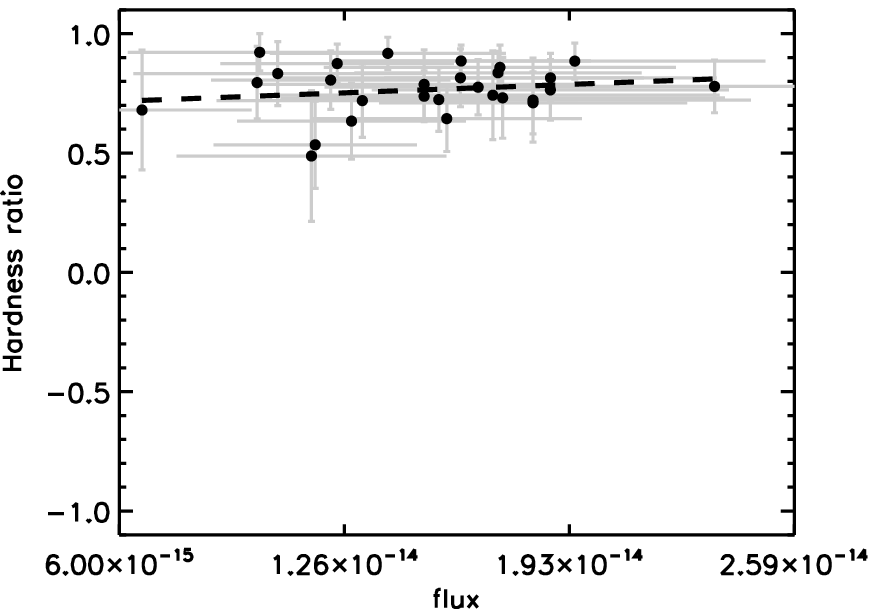}
\includegraphics[width=6cm]{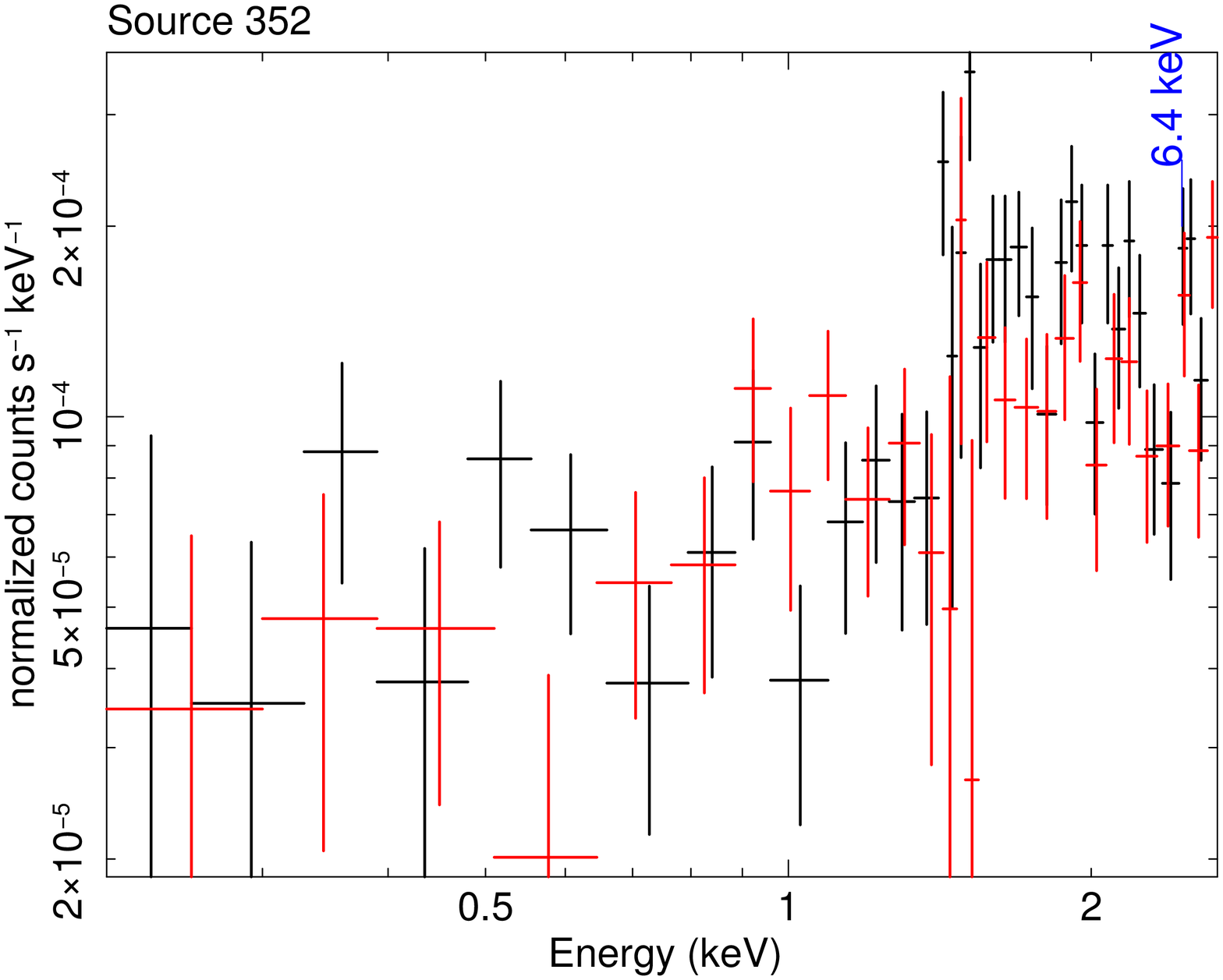}\\
\caption{Source 352.  Plots as in Fig. 2.
 }
         \label{Fig352fluxes}
   \end{figure*}

 \begin{acknowledgements}
   This work is based on observations by
XMM-Newton
, an ESA science mission
with  instruments  and  contributions  directly  funded  by  ESA  member
states and NASA.
We acknowledge the referee for useful suggestions. 
   SF acknowledges the Swedish National Space Board and
the Knut \& Alice Wallenberg Foundation. SF also acknowledges the group of Particle and Astroparticle Physics and the students of the KTH for raising interesting questions about this work. 
      FJC acknowledges financial support through grant AYA2015-64346-C2-1P (MINECO/FEDER).
\end{acknowledgements}

\bibliographystyle{aa}
\bibliography{bibtex}

\end{document}